\documentclass[%
 reprint,
 amsmath,amssymb,
 aps,
]{revtex4-1}

\usepackage{graphicx}
\usepackage[rgb]{xcolor}
\usepackage{dcolumn}
\usepackage{bm}
\usepackage{amsmath}
\usepackage{amsthm}
\theoremstyle{definition}
\newtheorem{definition}{Def.}
\newcommand{\norm}[1]{\left\lVert#1\right\rVert}
\usepackage{nomencl}
\makenomenclature
\usepackage{listings}

\usepackage[hidelinks]{hyperref}

\begin{document}

\title{Diffusion geometry of multiplex and interdependent systems}

\author{Giulia Bertagnolli}
\email[Corresponding author:~]{giulia.bertagnolli@unitn.it}%
\affiliation{CoMuNe Lab, Fondazione Bruno Kessler, Via Sommarive 18, 38123 Povo (TN), Italy}
\affiliation{University of Trento, Department of Mathematica, Via Sommarive, 14, 38123 Povo (TN), Italy}

\author{Manlio De Domenico}%
\email[Corresponding author:~]{mdedomenico@fbk.eu}%
\affiliation{CoMuNe Lab, Fondazione Bruno Kessler, Via Sommarive 18, 38123 Povo (TN), Italy}

\begin{abstract}
Complex networks are characterized by latent geometries induced by their topology or by the dynamics on them.
In the latter case, different network-driven processes induce distinct geometric features that can be captured by adequate metrics.
Random walks, a proxy for a broad spectrum of processes, from simple contagion to metastable synchronization and consensus, have been recently used in [Phys. Rev. Lett. 118, 168301 (2017)] to define the class of diffusion geometries and pinpoint the functional meso-scale organization of complex networks from a genuine geometric perspective.
Here, we firstly extend this class to families of distinct random walk dynamics -- including local and non-local information -- on multilayer networks -- a paradigm for biological, neural, social, transportation, and financial systems -- overcoming limitations such as the presence of isolated nodes and disconnected components, typical of real-world networks.
Secondly, we characterize the multilayer diffusion geometry of synthetic and empirical systems, highlighting the role played by different random search dynamics in shaping the geometric features of the corresponding diffusion manifolds.
\end{abstract}

\maketitle

\section{\label{sec:intro} Introduction}

Complex networks, an abstract representation of the structural and functional backbone of complex systems, exhibit a wide spectrum of geometric features, from self-similarity to latent hidden metric spaces and topology~\cite{song2005self,song2006origins,serrano2008self,taylor2015topological}, which have been successfully exploited to gain new insights about the {structure} and dynamics of social~\cite{papadopoulos2012popularity}, neural~\cite{allard2020navigable}, transportation~\cite{boguna2009navigability,xu2020modular} and communication systems~\cite{boguna2010sustaining}, to mention a few emblematic examples (see Ref.~\cite{Boguna2021} for a review).
More recently, it has been shown that even the dynamics on complex networks can induce complex geometries which cannot be understood from inspecting only structural ones.
Such geometries have been studied in the case of a broad class of spreading phenomena: communicability distance has been introduced to study information exchange based on walks~\cite{estrada2012communicability}; an effective distance has been introduced for contagion processes~\cite{brockmann2013hidden}; a diffusion distance has been introduced to study collective phenomena such as synchronization, consensus and random searches~\cite{de2017diffusion}; a temporal distance has been defined to study how {perturbations} spread in biological systems~\cite{hens2019spatiotemporal}.
The common rationale is to model the propagation of information with network dynamics and investigate the {induced distances} between pairs of nodes.

Random walks are emblematic examples used for modeling diffusion and transport dynamics from lattices to disordered media and quantum systems, across more than a century~\cite{kac1947random,bouchaud1990anomalous,havlin2002diffusion,whitfield2010quantum,venegas2012quantum,zaburdaev2015levy,giuggioli2020exact}.
They provide both an intuitive -- and often analytically treatable -- mathematical framework and a rich physical model that can be used to map a wide spectrum of random processes and observed features, especially in the contest of complex networks, where the interplay between the underlying topology and the dynamics on it is responsible for the system's function~\cite{masuda2017}.
The properties of the random walks reflect particular structural features of a system: for instance, they can be used to unravel the meso-scale organization of systems from a functional perspective~\cite{rosvall2008maps,delvenne2010stability,faccin2014community,lambiotte2014random,de2015identifying};
localization effects can be observed around topological defects~\cite{burda2009localization}; the average number of different sites visited in a time interval by a random walker, the so-called coverage, can be used to quantify the system resilience to random failures~\cite{de2014navigability}; the relative importance of system's units can be quantified in terms of their ability to attract the overall flow, encoded by random walkers~\cite{brin1998anatomy};
the dynamical features of the information flowing through a network can be understood~\cite{de2016spectral}, and even enhanced~\cite{Ghavasieh2020}, in terms of the spectral entropy defined by statistical features of random walks.

Based on these dynamics, one can also define similarity measures that reflect the ability of the units to exchange information~\cite{coifman2006diffusion,cao2013going,hammond2013graph,de2017diffusion}.
In~\cite{de2017diffusion} diffusion distance was defined for single-layer networks and it has proven to be useful for characterizing the functional structure of complex networks, e.g., identifying functional clusters, or central nodes~\cite{bertagnolli2019}.
It also provides a {real}-valued distance function which, in applications where the global information about the underlying topology is missing or partial, is more informative than the purely topological length of shortest-paths.

\begin{figure}[!h]
  \centering
  \includegraphics[width=.45\textwidth]{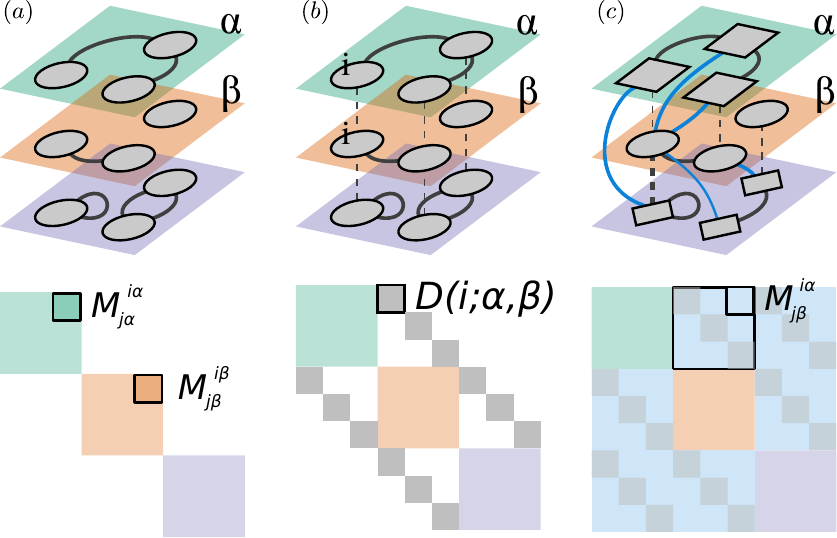}
  \caption{Different multilayer networks and their supra-adjacency matrix representation~\cite{gomez2013diffusion,de2013mathematical}: (a) an edge-colored multigraph, with layers corresponding to colors and no inter-layer connections; (b) a multiplex network where the replicas in the different layers are interconnected sequentially, a type of intertwining or diagonal coupling, and (c) the most general interconnected case, where inter-layer connections are not restricted to replicas (exogenous interactions).
Latin letters denote nodes, while Greek letters are used for layers.
{$D(i;\alpha \beta)$ is the intensity of the intertwining between state nodes $i, \alpha$ and $i, \beta$.
$M^{i \alpha}_{j \beta}$ describes the general inter-layer connectivity.}}
  \label{fig:multilayers}
\end{figure}

However, in many real systems entities interact in multiple ways; think, for example, to locations connected through different transportation modes or to metabolites in a biological network linked by various types of chemical reactions.
The evolution of the traditional graph-theoretical tools allows us, nowadays, to represent this multifaceted information by means of layers defining multilayer networks~\cite{de2013mathematical,Kivela2014}.
 Figure~\ref{fig:multilayers} shows distinct representations of the multiplexity and the interdependency observed in empirical complex networks.
From a mathematical perspective, taking into consideration the variety of connectivity patterns and coupling among layers and dynamics~\cite{de2016physics} requires multilinear algebra and tensors~\cite{de2013mathematical} to be efficiently represented and treated analytically.

In this study our contribution is twofold: on the one hand, we generalize multilayer random walk dynamics for the analysis of more realistic empirical systems, e.g., allowing for the existence of isolated nodes and components; on the other hand, we extend the framework of diffusion geometry~\cite{de2017diffusion}{, that is, the study of the shape of networks embedded into space through diffusion dynamics, to the realm} of multilayer systems, highlighting the dependence of the diffusion space on the interplay between structure and dynamics within and across layers.

The paper is organized as follows: in the next section we recall the basic facts on random walks and diffusion processes on {single-layer} networks and then generalize them to the multilayer framework.
In Sec.~\ref{sec:diffudist} we extend the definition of diffusion distance to multilayer networks, presenting results obtained from the analysis of synthetic models.
We conclude with applications to empirical social and transportation systems in Sec.~\ref{sec:applications}, followed by Discussion and Conclusions.

\section{Diffusion Dynamics in Complex Networks}\label{sec:diffusion}

In a group of individuals, a piece of news spreads differently if everyone knows each other or if there are subgroups of individuals that do not communicate with the others: the structure of this social network influences the diffusion of information.
Similarly, one observes a strong influence of the topology on the dynamics of a broad class of empirical systems.
Because of this influence of the structure on the evolution of dynamical processes, the latter are often used as a proxy for probing and uncovering structural properties of complex networks.
A simple strategy to explore a network is to start from a node, to choose at random an (outgoing) edge and to follow it toward the next node, i.e., to perform a random walk on the network with transition probabilities prescribed by the network connectivity.
Random walks (RWs) represent a useful model for diffusion processes.
Here we will focus on node-centric continuous-time random walks (CTRW)\cite{masuda2017}, continuous-time Markov chains (MC) on the vertex set of a network with transition rates depending on the network structure and specific navigation rules.

\subsection{Random walks and diffusion distance on single-layer networks}\label{ssec:monoplex}

Let us consider a weighted directed network $G=(V, E)$ without isolated nodes and with (possibly weighted) adjacency matrix $\mathbf{W} = (W_{ij}) \in \mathbb{R}^{N \times N}$, and two nodes $i, j\in V$.
In a discrete {classical} random walk the transition probability from $i$ to $j$ in one time step is given by $p_{ij} = \frac{W_{ij}}{s_i}$; {the probability of the random walker being in node $j$ after $n$ steps is encoded in the $j-$th component of the row vector $\mathbf{p}(n)$ and can be found by means of the equation $\mathbf{p}(n) = \mathbf{p}(n-1)\mathbf{D}^{-1}\mathbf{W}$, where $\mathbf{D}$ is the diagonal matrix of out-strengths $D_{ii} = s_i =\sum\limits_{j=1}^N W_{ij}$.}
The continuous-time random walk corresponding to this jump chain is described by the forward equation
{
\begin{equation}\label{eq:CTRW-ode}
\dot{\mathbf{p}}(t) = -\mathbf{p}(t) \tilde{\mathbf{L}}\\
\end{equation}}
where $\tilde{\mathbf{L}} = \mathbf{I}-\mathbf{D}^{-1} \mathbf{W}$ is the random walk normalized Laplacian and $-\tilde{\mathbf{L}}$ is the generator of the continuous-time Markov chain; see Appx.~\ref{appx:rws} for further details.
The transition probabilities {after time $t$} are given by the solution of \eqref{eq:CTRW-ode} with {some initial condition $\mathbf{p}(0) = \mathbf{p}_0$}, i.e., $p_{ij}(t) = (\mathbf{p}_0 e^{-t\tilde{\mathbf{L}}})_{ij}$.
Rewriting \eqref{eq:CTRW-ode} as a system with $\mathbf{P}(t)$ being a matrix and with the initial distribution given by the identity matrix, $\mathbf{P}(0) = \mathbf{I}$, we obtain the unique solution $\mathbf{P}(t) = e^{-t\tilde{\mathbf{L}}}$.
Its $i-$th row $\left(e^{-t\tilde{\mathbf{L}}}\right)_{i\cdot} = \mathbf{p}(t | i)$ is the probability vector corresponding to $\mathbf{p}_0 = \mathbf{e}_i$ and
$\left(e^{-t\tilde{\mathbf{L}}}\right)_{ij}$ is the probability of being in $j$, after time $t$, starting in $i$ with probability 1.
The diffusion distance~\cite{de2017diffusion} is the $L^2-$norm of the difference between rows of the matrix $e^{-t\tilde{\mathbf{L}}}${\textemdash the difference between the two posterior distributions at time $t$ obtained starting from $i$ and $j$ respectively, that is}
\begin{equation}\label{eq:diffu-dist-monoplex}
  D_t(i, j) = \norm{\mathbf{p}(t | i) - \mathbf{p}(t | j)}_2.
\end{equation}
{At each fixed time $t \geq 0$, $D_t(i, j)$ defines a distance on the vertex set, which is bounded in $[0, \sqrt{2}]$ for all pairs of nodes, indeed, $\norm{\mathbf{p}(t|i)}_2 \leq \norm{\mathbf{p}(t|i)}_1 = 1$.
Furthermore, two nodes $i \neq j$ are close w.r.t. $D_t$ if there is a large probability that two random walkers starting in $i$ and $j$ respectively, can be found at the same node at time $t$~\cite{coifman2006diffusion}, i.e., if the joint probabilities $p_{k}(t \mid i) p_{k}(t \mid j)$ of two (independent) Markov chains being in the same state $k$ at time $t$ are large.
See Appx.~\ref{appx:diffu-dist} for details.}
{Furthermore, as the diffusion time $t$ increases, and assuming that the walk is ergodic, $\mathbf{p}(t|i)$ will tend to the stationary distribution $\boldsymbol{\pi}$ and $D_{t}(i, j) \to 0$.}
{The family of diffusion distances define a family of kinematic geometries based on a network-driven process~\cite{Boguna2021}.
It has also been shown~\cite{de2017diffusion} that $t$ plays the role of a scale parameter and that the most persistent meso-scale structure of the network can be obtained averaging $(D_t)_{t>0}$ over time.
The average diffusion distance will be indicated by $\bar{D}_t$.}

{\begin{definition}[Diffusion space - 1]\label{def:ml-diffu-space}
The vertex set endowed with the diffusion distance, for a fixed value of $t$, is a metric space $(V, D_t)$.
Furthermore, a random walk dynamics maps $V$ to a set of points in space.
Indeed, given $i \in V$, $\left(e^{-t\tilde{\mathbf{L}}}\right)_{i\cdot} = \mathbf{p}(t | i)$ is a vector in $R^N$ and the diffusion distance at time $t$ between a pair of nodes $i, j$ is the Euclidean distance between the corresponding probability vectors.
We call this embedding of the network nodes in $\mathbb{R}^N$ trough the diffusion distance, the \textit{diffusion space} or \textit{diffusion manifold}.
\end{definition}}

On a given network, which is completely characterized by its adjacency matrix $\mathbf{W}$, we can define random walks with different flavors.
This enables us, not only to model a wider range of physical spreading processes, but also to remove the rather restrictive assumption of connectedness, which is necessary for writing the transition matrix $\mathbf{D}^{-1}\mathbf{W}$ of the classical random walk.
Consequently, also the family of diffusion distances can be generalized to different types of random walks, as well as to different types of networked systems.
In the remainder of this section, we extend the diffusion distance to multilayer networks, additionally exploring the varying patterns induced by different random walks dynamics~\cite{de2014navigability,de2015identifying}.
Of course, there are many other types of random walks that are not mentioned here, e.g., multiplicative processes~\cite{havlin1988random} or correlated random walks~\cite{gillis1955correlated}, {which} are often defined with a specific motivation.
Nevertheless, our framework is very general and, given a transition matrix $\mathbf{T}$, can be effortlessly {extended} considering the continuous-time Markov chain having exponential holding times with rate one and $\mathbf{T}$ as jumping matrix, see Appx.~\ref{appx:rws}.

Before moving to the multilayer case, let us introduce the tensorial notation~\cite{de2013mathematical} for the monoplex $G$, which allows us to generalize the formalism to multilayer networks.
The adjacency matrix $\mathbf{W}$ can be seen as a rank-2 tensor.
In this paper, we will not use the covariant notation and the Einstein summation convention, so that $W^i_j$ denotes the component $(i,j)$ of the tensor.
The master equation \eqref{eq:CTRW-ode}{, and the corresponding initial condition,} can be re-written as
\begin{equation}\label{eq:CTRW-ode-tensor}
\dot{{p}}_{j}(t) = - \sum\limits_{i=1}^N \tilde{{L}}^{i}_{j} {p}_{i}(t), \qquad {p}_{i}(0) = q_i
\end{equation}
with $\tilde{{L}}^{i}_{j} = \delta^{i}_{j} - T^{i}_{j}$ indicating the component $(i,j)$ of the random walk normalized Laplacian tensor, $\delta^{i}_{j}$ the Kronecker delta and $T^{i}_{j}$ the component $(i,j)$ of the transition probability tensor.
\nomenclature{$W^i_j$}{$ij-$th element of the weighted adjacency matrix of a graph seen as a rank-2 tensor}

\subsection{Random Walks on edge-colored multigraphs}\label{ssec:edge-colored}

\begin{figure*}[!htb]
  \centering
  \includegraphics[width=\textwidth]{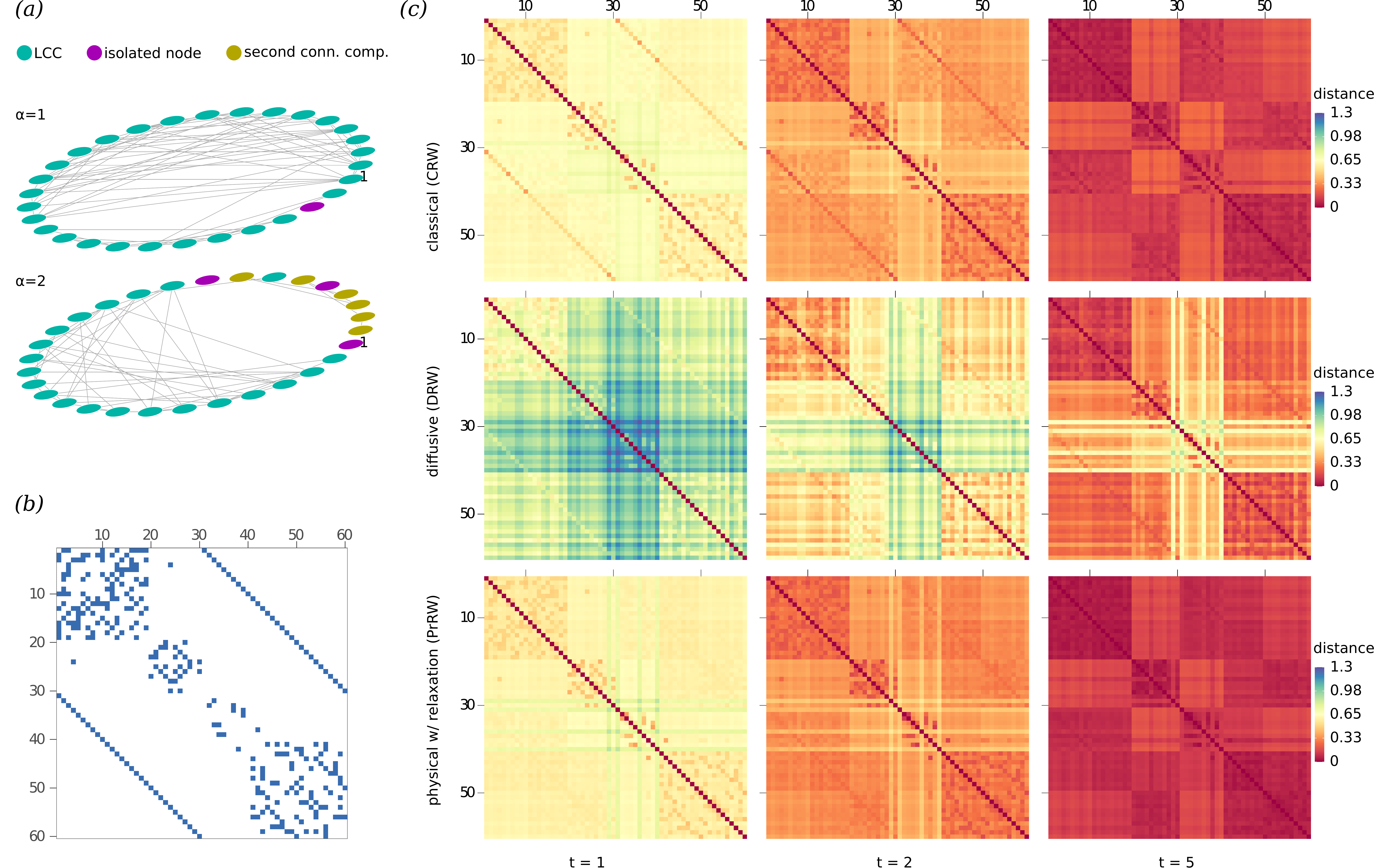}
  \caption{A synthetic {two-layer} network with unitary diagonal coupling $D(i; \alpha, \beta) = 1$, for all $i$ and $\alpha \neq \beta$, and its functional characterization. (a) Isolated nodes, {the largest connected component (LCC) and other} different components have been highlighted in the two layers. (b) The supra-adjacency matrix of the multiplex. (c) The $NL \times NL$ diffusion supra-distance matrices for three different random walk dynamics and $t = 1, 2, 5$. The PrRW is evaluated for $r=0.5$.}
  \label{fig:synth-ddms}
\end{figure*}

A multilayer network is defined by a set of $N$ nodes $V$
interacting with each other in multiple ways, simultaneously.
The different types of interaction can be encoded by colors and grouped into layers.
Each node $i \in V$ exists at least in one layer and, if it exists in multiple layers $\alpha \in \{1, \dots, L\}$, we usually {refer} to {$(i, \alpha)$} as a \textit{replica} or \textit{state node}, in contraposition to the physical node $i$.
Interactions of the same color determine the intra-layer connectivity, while the others are called inter-layer connections.
The simplest multilayer structure, depicted in Fig.~\ref{fig:multilayers}(a), is the edge-colored multigraph, where there is no additional information (e.g., no order relation) on the set of layers $\{1, \dots, L\}$ or, equivalently, inter-layer connectivity is not present.
{
\begin{definition}[Adjacency matrices for edge-colored multigraphs]\label{def:adjacency-matrix}
Given a edge-colored multigraph $G=(V, E, \{1, \dots, L\}, c)$, where $V$ is the set of physical nodes, $E$ is a multiset, $\{1, \dots, L\}$ is the set of colors, called layers from now on, and $c: E \to \{1, \dots, L\}$ is a surjective function assigning a color to each edge, we denote by $W^{i \alpha}_{j \alpha}$ the adjacency matrix in layer $\alpha = 1, \dots, L$.
\end{definition}
}
\nomenclature{$W^{i\alpha}_{j\alpha}$}{$ij-$th element of the weighted adjacency matrix of the layer $\alpha$ of a edge-colored multigraph. See Def.~\ref{def:adjacency-matrix}}
There are, essentially, two ways to define random walks on these networks. One possibility is to allow the random walker to follow a sequence of edges with different colors: colored edges are then treated as multiple edges and the degrees of a node counts all the edges, regardless of their colors~\cite{battiston2016}. This choice is equivalent to aggregate the multilayer system to a single layer and perform a classical random walk on it. This approach is not desirable in general, because one does not know \emph{a priori} if, and to which extent, information {is} lost while aggregating the structure of layers will affect the results.

The second approach, used successfully in other applications~\cite{Ghavasieh2020}, is to run independent dynamics on each layer and then integrate the transition matrices over the layers with the necessary normalization.
{Let us denote the transition probability from $i$ to $j$ in layer $\alpha = 1, \dots, L$ by $T^{i \alpha}_{j \alpha}$. Then the overall transition probability from $i$ to $j$ is obtained as follows:
\begin{equation}\label{eq:tmatrix-ec}
\langle T^i_{j} \rangle = \sum\limits_{\alpha = 1}^L m_{i, \alpha} T_{j \alpha}^{i \alpha}, \qquad \sum\limits_{\alpha=1}^L m_{i, \alpha} = 1 \quad \forall i \in V,
\end{equation}
where $m_{i, \alpha}\geq 0$ are weights enabling us to tune the importance of each layer, relative to node $i$ in the final dynamics of the system.
If $m_{i, \alpha} = \frac{1}{L}$ for all $i$ and $\alpha$, then $\langle T^i_{j} \rangle$ is simply the average transition matrix over all layers.
However, if we choose $m_{i, \alpha} = \frac{1}{\mu_i} \mathbf{1}_{\{s^{i \alpha}_{\alpha} \neq 0\}}$, with $s^{i \alpha}_{\alpha}$ being the out-strength of vertex $i$ in layer $\alpha$ and $\mu_i = \sum\limits_{\alpha=1}^L \mathbf{1}_{\{s^{i \alpha}_{\alpha} \neq 0\}}$ the multiplicity of node $i$, i.e., the number of layers in which $i$ is not isolated, then we can discard the effect of $i$ being isolated in one or more layers}.

This approach is more desirable than the first one, since it preserves more information related to diversity of connectivity patterns across layers.
The main difference between the two approaches emerges when the edges are weighted, besides colored: \emph{a priori} the scale (nominal, ordinal, ratio etc.) of the weights could be different and an inattentive summation could lead to errors.
In Eq.~\eqref{eq:tmatrix-ec}, instead, for each $i \in V$, we are taking a finite mixture of probability mass functions {with weights $m_{i, \alpha}\geq 0$ such that $\sum\limits_{\alpha=1}^L m_{i, \alpha} = 1$}; nothing prevents one to weight layers differently, if additional information is available. In the following, we consider equal importance for all layers {and choose $m_{i, \alpha} = \frac{1}{\mu_i} \mathbf{1}_{\{s^{i \alpha}_{\alpha} \neq 0\}}$}.
{
\begin{definition}[Transition matrix]\label{def:transition-matrix}
Given a set $V$ of $N$ vertices a \emph{transition matrix} $\mathbf{T}$ is an $N \times N$ matrix, whose entries $T^i_j$ are non-negative reals representing the probability of moving from state (node, in the realm of graphs) $i$ to $j$. $\mathbf{T}$ should also satisfy $\sum\limits_{j=1}^N T^i_j = 1$.
The $i-$th row of the matrix $\mathbf{T}$ is then a probability vector, whose $j-$th component is the probability of being in $j$ given that the previous state was $i$.
Given an edge-colored multigraph with state nodes $\{(i, \alpha): i=1, \dots, N \text{ and } \alpha = 1, \dots, L\}$ we define $T^{i \alpha}_{j \alpha}$ as the transition probability from $(i, \alpha)$ to $(j, \alpha)$. Transitions from $(i, \alpha)$ to $(j, \beta)$ are not considered here, since there is no inter-layer connectivity for edge-colored multigraphs and, until now we just have a collection of $L$, $N \times N$ transition matrices.
\end{definition}}

We now take advantage of the more involved notation of Sec.~\ref{ssec:monoplex}, to introduce multilayer interconnected networks.

\nomenclature{$T^i_j$}{The transition probability from $i$ to $j$ and $ij-$th element of the transition matrix on a graph seen as a rank-2 tensor. See Def.~\ref{def:transition-matrix}}
\nomenclature{$T^{i\alpha}_{j\alpha}$}{The transition probability from state node $(i, \alpha)$ to $(j, \alpha)$ in a edge-colored multigraph. It can be seen as the $ij-$th element of the transition matrix in layer $\alpha$. See Def.~\ref{def:transition-matrix}}

\subsection{Random walks on multilayer networks}

\begin{table*}[!htb]
\begingroup
\setlength{\tabcolsep}{8pt}
\renewcommand{\arraystretch}{2.5}
\begin{tabular}{l|c|c|c|c|r}
& CRW~\cite{de2014navigability} & PRRW~\cite{de2015ranking} & DRW~\cite{de2014navigability} & MERW~\cite{de2014navigability} & PrRW~\cite{de2015identifying} \\ \hline
$T_{j \beta}^{j \beta}$  & $\frac{M_{j \beta}^{j \beta}}{S^{j \beta}}$   & $r \frac{M_{j \beta}^{j \beta}}{S^{j \beta}} + (1 - r) \frac{1}{NL}$   & $\frac{s_{\max }+M_{j \beta}^{j \beta} - S^{j \beta}}{s_{\max }}$ & $\frac{M_{j \beta}^{j \beta}}{\lambda_{\max }}$ & $(1-r) \frac{M_{j \beta}^{j \beta}}{s^{j \beta}_{\beta}} + r \frac{M_{j \beta}^{j \beta}}{s^j}$ \\
$T_{j \beta}^{j \alpha}$ & $\frac{M_{j \beta}^{j \alpha}}{S^{j \alpha}}$ & $r \frac{M_{j \beta}^{j \alpha}}{S^{j \alpha}} + (1 - r) \frac{1}{NL}$ & $\frac{M_{j \beta}^{j \alpha}}{s_{\max }}$ & $\frac{M_{j \beta}^{j \alpha}}{\lambda_{\max }} \frac{V_{j\beta}}{V_{j\alpha}}$ & $r \frac{M_{j \beta}^{j \beta}}{s^j}$ \\
$T_{j \beta}^{i \beta}$  & $\frac{M_{j \beta}^{i \beta}}{S^{i \beta}}$   & $r \frac{M_{j \beta}^{i \beta}}{S^{i \beta}} + (1 - r) \frac{1}{NL}$   & $\frac{M_{j \beta}^{i \beta}}{s_{\max }}$ & $\frac{M_{j \beta}^{i \beta}}{\lambda_{\max }} \frac{V_{j \beta}}{V_{i \beta}}$   & $(1-r) \frac{M_{j \beta}^{i \beta}}{s^{i \beta}_{\beta}} + r \frac{M_{j \beta}^{i \beta}}{s^i}$ \\
$T_{j \beta}^{i \alpha}$ & $\frac{M_{j \beta}^{i \alpha}}{S^{i \alpha}}$ & $r \frac{M_{j \beta}^{i \alpha}}{S^{i \alpha}} + (1 - r) \frac{1}{NL}$ & $\frac{M_{j \beta}^{i \alpha}}{s_{\max }}$ & $\frac{M_{j \beta}^{i \alpha}}{\lambda_{\max }} \frac{V_{j \beta}}{V_{i \alpha}}$ & $r \frac{M_{j \beta}^{i \beta}}{s^i}$ \\
\end{tabular}
\endgroup
\caption{Transition probabilities for different random walks. (CRW) classical, (PRRW) PageRank, (DRW) diffusive, (MERW) maximal-entropy, and (PrRW) physical with relaxation random walks. {$s_{\max }=\max\limits_{i, \alpha}\left\{S^{i \alpha} \right\}$}; the jumping parameter of the PageRank RW is commonly indicated by $\alpha$, to avoid ambiguity, we indicate it by $r$; $\lambda_{\max }$ is the largest eigenvalue of the adjacency tensor and $V$ is its corresponding eigentensor, satisfying $\sum\limits_{i, \alpha} M_{j \beta}^{i \alpha} V_{i \alpha} = \lambda_{\max } V_{j \beta}$ (see~\cite{de2013mathematical} for details). {Note that for CRW, DRW and MERW, these transition rules generalize the one introduced in~\cite{de2014navigability} for the analysis of multiplex networks. The PrRW is defined as in~\cite{de2015identifying}, while PRRW generalizes the walk introduced in~\cite{de2015ranking}.}}
\label{tab:rws}
\end{table*}

\nomenclature{$M^{i\alpha}_{j\beta}$}{Element of the rank-4 adjacency tensor of a multilayer network}
\nomenclature{$s^{i \alpha}_{\alpha},~s^i,~S^{i \alpha}$}{Respectively: out-strength of the state node $(i, \alpha)$; multilayer out-strength $\sum\limits_\alpha s^{i \alpha}_{\alpha}$ of the physical node $i$ without inter-layer connections; out-strength of state node $i$ in layer $\alpha$, with inter-layer edges. The strength of node $i$ accounting solely for the inter-layer connections is obtained as $S^{i \alpha} - s^{i \alpha}_{\alpha}$.}

In the same way a monoplex can be represented by a rank-2 tensor, a multilayer network is completely characterized by its rank-4 adjacency tensor~\cite{de2013mathematical}. Let us indicate by $M_{j \beta}^{i \alpha}$ the components of the (weighted) multilayer adjacency tensor.
In this index notation $M_{j \beta}^{i \alpha}$ indicates the interaction between $i$ and $j$ in the same layer, while $M_{i \beta}^{i \alpha}$ denotes the strength of the intertwining of the replicas of $i$ in two distinct layers.
Depending on the inter-layer connectivity we can have different types of multilayers, as shown in Fig.~\ref{fig:multilayers}.
When inter-layer connections occur only between replicas of the same physical node, i.e., $M_{j\beta}^{i \alpha} = 0$ for $i\neq j$, the network is called a multiplex.
The term $M_{i\beta}^{i \alpha}$ encodes the intertwining between two replicas; it is a scalar depending, in general, on $i, \alpha, \beta$ and we indicate it by $D(i; \alpha, \beta)$, {to emphasizes the ``diagonal coupling''~\cite{Kivela2014,de2014navigability,masuda2017} between replicas.}
For the ease of visualization, we flatten the adjacency tensor into the so-called supra-adjacency matrix~\cite{kolda2009,gomez2013diffusion,de2013mathematical}, of dimension $NL \times NL$.
Its diagonal blocks are the adjacency matrices of the monoplexes, while its off-diagonal blocks contain the information on the inter-layer connectivity.

We henceforth indicate by {$s^{i \alpha}_{\alpha} = \sum\limits_{j=1}^{N} M^{i \alpha}_{j \alpha}$}, the out-strength of node $i$ in layer $\alpha$ and by {$s^{i} = \sum\limits_{\alpha=1}^L s^{i \alpha}_{\alpha} = \sum\limits_{\alpha} \sum\limits_{j} M_{j \alpha}^{i \alpha}$} its multilayer out-strength, discarding the inter-layer edges {; while, if inter-layer edges are accounted for, $S^{i \alpha}=\sum\limits_{j} \sum\limits_{\beta} M^{i \alpha}_{j \beta}$ denotes the out-strength of node $i$ in layer $\alpha$.}
Observe that the inter-layer strengths are then obtained as {$S^{i \alpha} - s^{i \alpha}_{\alpha}$.}

Generally, the presence of isolated nodes or components, as well as a heterogeneous or mixed distribution of inter-layer connectivity is discarded to facilitate the analytical framework. Conversely, here we do not force any assumption on the structure of the multilayer system: if some units are not present in all layer -- which is often the case for real data -- we will add artificial replicas, which will appear as isolates and we will account for them adequately.
We henceforth indicate by $V$ the common vertex set and by $N = |V|$ the number of physical nodes.
Although the isolated nodes do not modify the connectivity of the connected component of the layer, they could be troublesome for the evaluation of transition probabilities.
We have to distinguish three cases (i) {$S^{i \alpha}-s^{i \alpha}_{\alpha} = 0$,} (ii) $s^{i \alpha}_{\alpha} = 0$, and (iii) $s^i = 0$. The latter corresponds to the trivial case, where node $i$ is isolated in every layer, so that $i$ can be simply removed from the vertex set.
We can assume, without loss of generality, {that} $s^i > 0$ for all $i \in V$.
(i) and (ii) -- corresponding to no inter-layer and no intra-layer connections, respectively -- constitute a problem only if {$S^{i \alpha} = 0$} for some $\alpha$.
In this case one could see $(i, \alpha)$ as an absorbing state with the probability of remaining there equal to 1; another option is to teleport the random walker in $(i, \alpha)$ to any $(j, \beta)$ with uniform probability $\frac{1}{NL}$. As for other approaches~\cite{brin1998anatomy,rosvall2008maps,de2015identifying}, we opted for the second option, since it decreases the occupation probability of the state node $(i, \alpha)$.

In the multilayer framework the probability transitions in one time step constitute the components $T_{j \beta}^{i \alpha}$ of a rank-4 tensor and we can expand the RW equation to highlight different contributions of jumps and switches in the dynamics
\begin{align*}
  p_{j \beta}(t + 1) =
  & \underbrace{T_{j \beta}^{j \beta} p_{j \beta}(t)}_{\text{stay}} +
  \underbrace{\sum_{\alpha = 1 \atop \alpha \neq \beta }^L T_{j \beta}^{j \alpha} p_{j \alpha}(t)}_{\text{switch}} + \\
  & +
  \underbrace{\sum_{i=1 \atop i \neq j}^{N} T_{j \beta}^{i \beta} p_{i \beta}(t)}_{\text{jump}} +
  \underbrace{\sum_{\alpha=1 \atop \alpha \neq \beta}^{L} \sum_{i=1 \atop i \neq j}^{N} T_{j \beta}^{i \alpha} p_{i \alpha}(t)}_{\text{switch and jump}}
\end{align*}

Its continuous-time version is described by the forward equation
\begin{equation}\label{eq:CTRW-ml-ode-tensor}
  \dot{p}_{j \beta}(t) = -\sum\limits_{i, \alpha} \tilde{L}_{j \beta}^{i \alpha} p_{i \alpha}(t)
\end{equation}
where $\tilde{L}_{j \beta}^{i \alpha}=\delta_{j \beta}^{i \alpha}-T_{j \beta}^{i \alpha}$ {and ${p}_{j \beta}(t)$ is the component of a rank-2 tensor indicating the probability of finding a random walker in node $j$ of layer $\beta$ at time $t$. When we want to highlight the initial condition, e.g., ${p}_{i \alpha}(0)=1$ for the RW dynamics, we will use the conditional notation ${p}_{j \beta}(t|(i, \alpha))$.}

On a given network structure we can define different types of random walks, depending on the specific rules we want our random walker to explore the network. For instance, in PageRank~\cite{brin1998anatomy} a \textit{teleportation} or \textit{jumping} parameter gives the possibility to the walker to reach also nodes that are not directly connected to the current node.
The flavors of the random walks are given by their transition rules, which depend on the structure of the network, as a function of the adjacency tensor.
The RW presented in the monoplex case is referred to as a classical random walk (CRW).
Additionally to the classical random walk, we look here at a family of diffusion distances based on four other random walk types: multilayer PageRank (PRRW) generalizing Refs.~\cite{brin1998anatomy,de2015ranking}, multilayer diffusive (DRW) generalizing the one defined in Ref.~\cite{de2014navigability}, maximal-entropy (MERW) generalizing Refs.~\cite{burda2009localization,de2014navigability}, and physical random walk with relaxation (PrRW)~\cite{de2015identifying}, whose transition probabilities are shown in Tab.~\ref{tab:rws}.

{\paragraph{Classical (CRW), PageRank (PRRW), maximal-entropy (MERW) random walks.} A classical random walker in a state node $(i, \alpha)$ of a multilayer network can move according to intra- or inter-layer connections, with uniform probability.
The normalizing factor for the transition probabilities is then the total strength $S^{i \alpha}$.
The PageRank RW differs from the classical one in that the walker moves with probability $r$ according to its edges (of both types) and with probability $1-r$ it may jump (or teleport itself) to any state node in the network.
In maximal-entropy random walk, as it can be seen in Tab.~\ref{tab:rws}, the transition probabilities are governed by the largest eigenvalue, and corresponding eigenvector, of the adjacency tensor. Here we assume the networks to be undirected. In the discrete-time framework, it has been shown~\cite{burda2009localization} that all the trajectories of length $l$ between two given nodes have the same probability and that the MERW has maximal entropy production rate among all walks on the network.}

{\paragraph{Diffusive random walk (DRW).} The jumping process described by the transition probabilities in DRW~Tab.~\ref{tab:rws} is characterized by a non-zero probability of remaining in a given state node $(i, \beta)$, $1-\frac{S^{i \beta}}{s_{\max}}$, and the probability of following out-going links with a probability normalized over the maximum strength over all nodes, $s_{\max}$.}

{\paragraph{Physical random walk with relaxation (PrRW).} The} physical random walk has been defined in~\cite{de2014navigability} to describe those dynamics where the state nodes have a ``common memory'', so that the information diffuses instantaneously across replicas.
Think, for instance, of the system of virtual interactions among individuals, who may have a profile (an alter-ego) in different social networks. A person can then exchange information in a particular social network using the (intra-layer) connections of its alter-ego in that social system, but she/he has always a complete knowledge of the information across the layers.
In this case, inter-layer connections between replicas of different physical nodes have no physical meaning and, consequently, are ignored.
The physical random walk with relaxation (PrRW)~\cite{de2015identifying} is a variant on the physical random walk, where the assumption on the complete knowledge of intertwining between layers is dropped. It can be seen from Tab.~\ref{tab:rws} that its transition probabilities contain a trade-off between intra- and inter-links, which are followed with probability $1-r$ and $r$ respectively.
If not differently stated, we consider here $r = 0.5$.

In the following section we will define a multilayer distance which reflects how nodes exchange information in and between the layers.

\section{Diffusion Distance in Multilayer Systems}\label{sec:diffudist}

\begin{figure}[!htb]
  \centering
  \includegraphics[width=.35\textwidth]{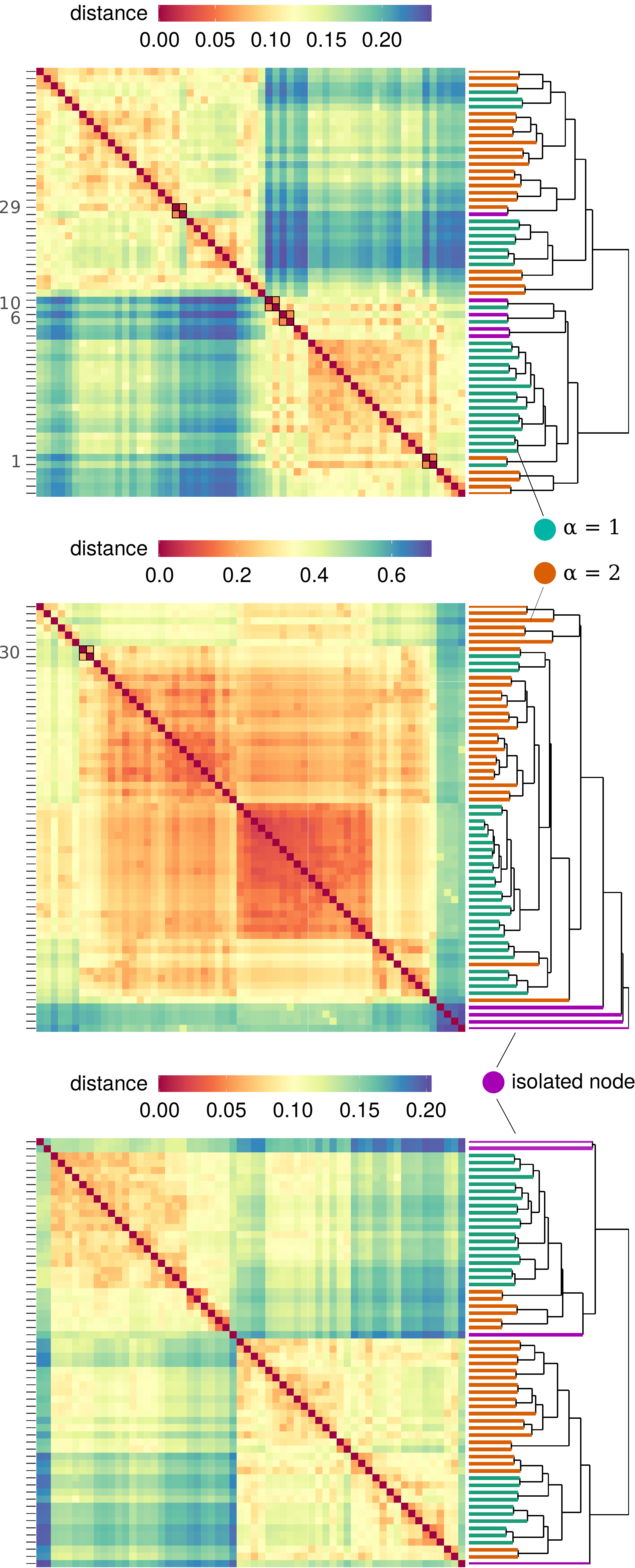}
  \caption{Average diffusion {distance} matrices of the {two-layer} synthetic network shown in Fig.~\ref{fig:synth-ddms}(a), w.r.t. three RW dynamics: (top) classical, (middle) diffusive, and (bottom) physical with relaxation $r=0.5$. The state nodes $(i, \alpha)$ are colored in the corresponding dendrogram according to the layer they belong to, except for isolated nodes, highlighted in purple. Small squares with a black border indicate those state nodes grouped together, e.g., $(29, 1), (29, 2)$ in the top panel.}
  \label{fig:synth-avg-hclu}
\end{figure}

In the multilayer framework the probability of finding a random walker at a given node and layer is encoded in a time-dependent tensor,  whose component $p_{j \beta}(t)$, corresponds to the probability of finding the random walker in the state $(j, \beta)$, for a given initial distribution.
Similarly to \eqref{eq:diffu-dist-monoplex}, we can then define the diffusion distance between state nodes $(i, \alpha)$ and $(j, \beta)$ as

\begin{equation}\label{eq:diffu-dist-ml}
 D_{t}^2((i, \alpha), (j, \beta))= \sum\limits_{k, \gamma}(p_{k\gamma}(t|(i, \alpha)) - p_{k\gamma}(t|(j, \beta)))^2.
\end{equation}
{As for the monoplex case, we can write the transition probabilities in terms of the exponential of the supra-Laplacian matrix,
so that Eq.~\eqref{eq:diffu-dist-ml} can be written as
\begin{equation}\label{eq:diffu-dist-ml-2}
    D_{t}^2((i, \alpha), (j, \beta)) = \sum\limits_{k, \gamma}\left[\left(e^{-t \tilde{L}}\right)^{i \alpha}_{k \gamma} - \left(e^{-t \tilde{L}}\right)^{j \beta}_{k \gamma}\right]^2
\end{equation}
and can be characterized by the spectrum of the Laplacian tensor~\cite{de2013mathematical,gomez2013diffusion}.
At $t$ fixed, $D^{i \alpha}_{j \beta}(t)=D_{t}((i, \alpha), (j, \beta))$ is the component of a metric tensor, which can be flattened into a $NL \times NL$ matrix indicated by $\mathbf{D}_{t}$ and called the supra-distance matrix, using the same nomenclature of ``supra-matrices'' introduced in~\cite{Kivela2014}.}
{Let us now summarize the just introduced concepts:
\begin{definition}[Diffusion distance, tensor, supra-distance matrix, average diffusion distance and normalized supra-distance matrix.]\label{def:ml-diffu-dist}
Given a multilayer network and a random walk dynamics on it, whose transition probabilities at time $t$ are given by $p_{j \beta}(t|(i, \alpha)) = \left(e^{-t \tilde{L}}\right)^{i \alpha}_{j \beta}$, we define the diffusion distance between state nodes $(i, \alpha)$ and $(j, \beta)$ at time $t$, $D_{t}((i, \alpha), (j, \beta))$, as in Eq.~\eqref{eq:diffu-dist-ml-2}. The terms of the resulting diffusion distance tensor are denoted by $D^{i \alpha}_{j \beta}(t)=D_{t}((i, \alpha), (j, \beta))$. We denote by $\mathbf{D}_t$ the diffusion supra-distance matrix, resulting from the flattening of the rank-4 tensor. The average diffusion distance~\cite{de2017diffusion} is defined as
\begin{equation*}
    \bar{D}_t = \frac{1}{t_{\max}} \sum_{\tau = 1}^{t_{\max}} D_{\tau}
\end{equation*}
where $t_{\max}$ is a temporal cutoff.
The corresponding average diffusion supra-distance matrix is indicated by $\bar{\mathbf{D}}_t$. Finally, we introduce the normalized diffusion supra-distance matrix~\cite{de2017diffusion}
\begin{equation*}
    \tilde{\mathbf{D}}_t = \frac{\mathbf{D}_t}{\max \mathbf{D}_t}.
\end{equation*}
\end{definition}}
\nomenclature{$p_{k\gamma}(t|(i, \alpha))$}{probability of being in state node $(k, \gamma)$ at time $t$ given the initial condition $p_{i\alpha}(t=0)=1$; furthermore, it holds that $p_{k\gamma}(t|(i, \alpha)) = \left(e^{-t L}\right)^{i \alpha}_{k \gamma}$.}
\nomenclature{$D_{t}^2((i, \alpha), (j, \beta))$}{Diffusion distance between state nodes $(i, \alpha)$ and $(j, \beta)$ at time $t$; see Eq.~\eqref{eq:diffu-dist-ml}.}
\nomenclature{$D^{i \alpha}_{j \beta}(t)=D_{t}((i, \alpha), (j, \beta))$}{component of the diffusion metric tensor, at a fixed $t$. $\mathbf{D}_t$ indicates its flattened version, called the diffusion supra-distance matrix, see Def.~\ref{def:ml-diffu-dist}}
\nomenclature{$\bar{D}_t$}{Average diffusion distance, i.e., diffusion distance averaged over time, usually over $(0, t_{\max}]$ at discrete steps, see Def.~\ref{def:ml-diffu-dist}}
\nomenclature{$\bar{\mathbf{D}}_t$}{Average diffusion distance matrix, see Def.~\ref{def:ml-diffu-dist}}
\nomenclature{$\tilde{\mathbf{D}}_t$}{Normalized diffusion distance (supra-)matrix, see Def~\ref{def:ml-diffu-dist}}

\begin{figure*}[!htb]
  \centering
  \includegraphics[width=0.75\textwidth]{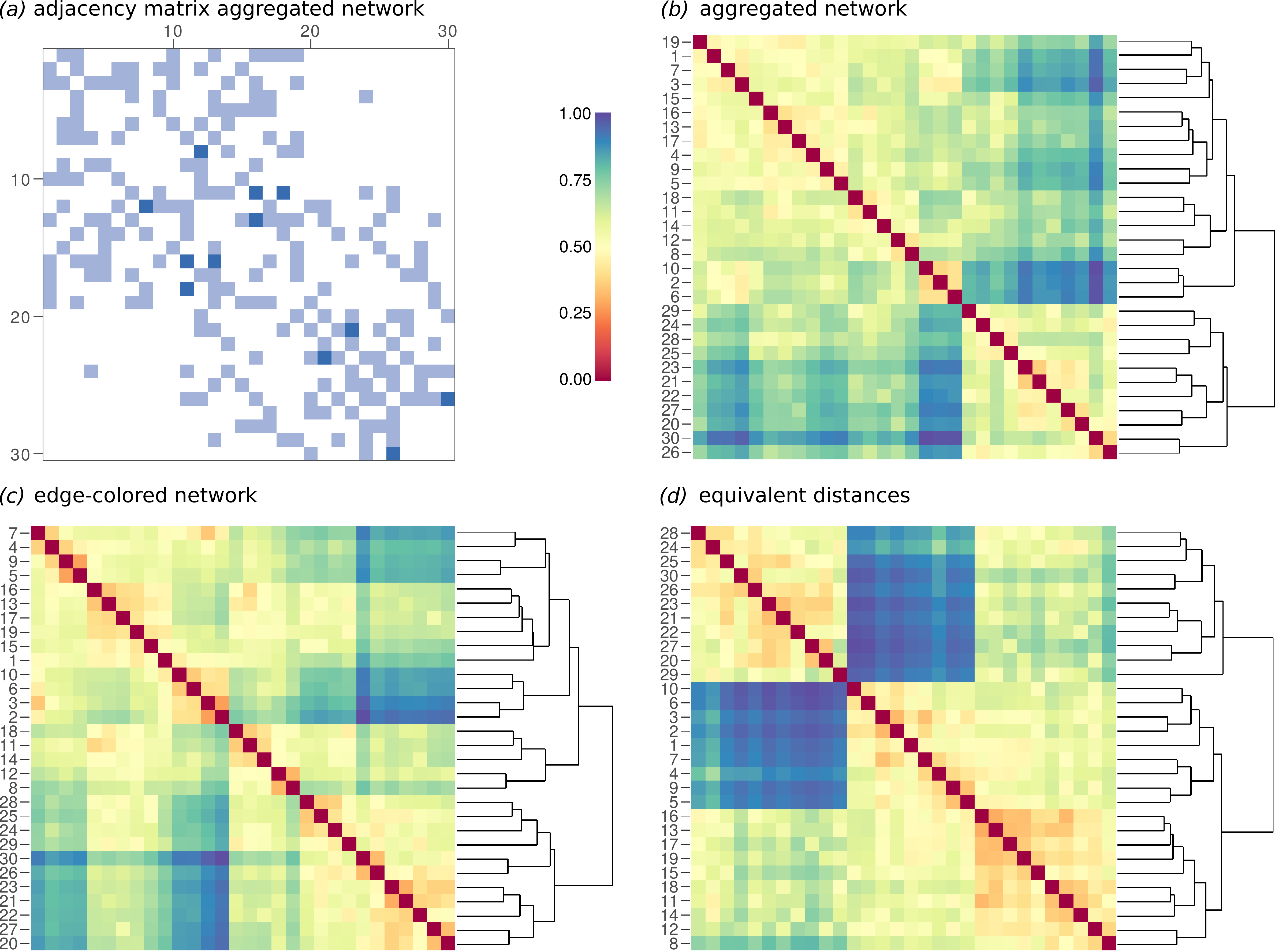}
  \caption{Comparing different types of {($N \times N$) average diffusion distance matrices between} physical nodes, regardless of layers. (a) The weighted adjacency matrix of the aggregated network. The diffusion distance matrix (w.r.t. the CRW and averaged over time) (b) of the aggregated network corresponding to the synthetic multiplex of Fig.~\ref{fig:synth-ddms}(a) and (c) of the edge-colored network. (d) The equivalent diffusion {distance matrix} obtained with the reduction of Eq.~\eqref{eq:equiv-ddm}. The distances are rescaled to $[0, 1]$, normalizing by the corresponding maximum {over all node pairs} (aggregated: $\max(\bar{\mathbf{D}}_t) \approx 0.13$, edge-colored: $\max(\bar{\mathbf{D}}_t) \approx 0.16$, equivalent distances: $\max(\bar{\mathbf{D}}_t^{\text{eq}}) \approx 0.11$).}
  \label{fig:synth-aggregation}
\end{figure*}

Figure~\ref{fig:synth-ddms} shows the supra-distance matrix $\mathbf{D}_{t}$ for the three more diverse RW dynamics on a synthetic multilayer network.
In each layer, we have a network with $N = 30$ nodes generated from a stochastic block model with two blocks. We chose the probabilities in order to have a diverse topology: dense groups, disconnected components and isolated nodes.
The coupling between the layers is $D(i; 1, 2) = D(i; 2, 1) = 1$, as shown in the supra-adjacency matrix of Fig.~\ref{fig:synth-ddms}(b). The columns of panel (c) represent different diffusion times $t$.
Recall that the diffusion time plays the role of a scale parameter~\cite{de2017diffusion} and that the continuous-time Markov chain has exponentially distributed holding times with rate $\lambda = 1$, i.e., the expected time occurring among each step of the RW is 1. $D_{t=1}$ is then a function of the micro-scale structure of the multiplex and here the isolated nodes are clearly visible.
Remarkably, the distances w.r.t. the diffusive random walk span a larger interval than the others and differentiate very effectively the subgroups in each layer.
This is a consequence of the diffusive dynamics, where the jumping probabilities do not depend on the vertex, but on the strength of hubs.
It is also worth noticing that, those nodes, which are disconnected from the largest connected component of the second layer, are generally nearer to the nodes in the first layer than to the nodes in their same layer.
To unveil the {persistent} meso-scale structures of the network we averaged the diffusion {supra-distance} matrices for up to $t_{max} = N$~\cite{de2017diffusion} and run a hierarchical clustering on the average diffusion supra-distance matrices, summarizing the results in Fig.~\ref{fig:synth-avg-hclu}.
As expected, the presence of inter-links moves loosely connected nodes {closer} to their replica in the other layer and the clustering does not separate the two layers.
{In this and the following plots, the state nodes in the supra-distance matrices are re-ordered according to the dendrogram returned by the complete linkage method for hierarchical clustering, (R) \textbf{hclust}~\cite{Rstats}. At the beginning, the two nearest nodes are grouped together, then at each step, the distance between the clusters is defined as the maximum distance between their nodes and the two nearest clusters are aggregated, until all the nodes are in the same group. In~\cite{de2017diffusion} it has been shown that the dichotomy between fast-shrinking (intra-community) and slow-shrinking (inter-communities) distances probes the meso-scale structure at different resolutions and that the most persistent and representative meso-scale structure is the one maximizing the average diffusion distance between clusters. Here, we use the dendrograms just as a guide for the eye and do not delve into the multilayer community detection~\cite{de2015identifying,edler2017mapping} or the analysis of the hierarchies~\cite{perotti2020}.}

Finally, we provide a grounded way to summarize the supra-distance matrix into an $N \times N$ matrix collecting the diffusion distances among the physical nodes, regardless of the layers.
The $ij-$elements of the diagonal blocks of the supra-matrix, $\left\{D_t((i, \alpha), (j, \alpha)), \alpha \in 1, \dots, L \right\}$, can be seen, using the jargon of electrical circuits, as resistances in parallel between the physical nodes $i, j$.
Their equivalent resistance can then be found through the parallel sum of the resistances between the replicas{, yielding to the equivalent diffusion distance:}
\begin{equation}\label{eq:equiv-ddm}
  D_t^{\text{eq}}(i, j) = \left(\sum_{\alpha=1}^L \frac{1}{D_t((i, \alpha), (j, \alpha))} \right)^{-1}.
\end{equation}
{Note the equivalent diffusion distance (briefly, \textit{equivalent distance}, if not ambiguous) is as an aggregation of} diffusion distances across layers, and it is not related to the concept of resistance distance~\cite{klein1993resistance}.
\nomenclature{$D_t^{\text{eq}}(i, j)$}{Equivalent diffusion distance between physical nodes, see Eq.~\eqref{eq:equiv-ddm}, and the corresponding distance matrix $\mathbf{D}_t^{\text{eq}}$}

The {corresponding} equivalent distance matrix, denoted by $\mathbf{D}_t^{\text{eq}}$, is quite different from the one obtained evaluating the diffusion distance on the aggregated network, as shown in Fig.~\ref{fig:synth-aggregation}.
Here the distance matrices have been rescaled in $[0, 1]$ by division through their respective maximum values, reported in the figure caption.
Looking back at Fig.~\ref{fig:synth-ddms}(a), we can see that the nodes from 1 to 19 are densely connected in layer 1, while node 8 and nodes from 11 to 30 form the largest connected component of the second layer. In both distance matrices there is a clear block corresponding to the last ten nodes. However, only the equivalent diffusion distance matrix captures the particular position of the first ten nodes (removing 8): in the multiplex they are distant from the last ten, because they belong to different communities in layer 1 and to disconnected components in layer 2.
We also compare the aggregated network distance matrix Fig.~\ref{fig:synth-aggregation}(b) to the one of the edge-colored multigraph obtained removing the inter-layer links from the multilayer network, shown in Fig.~\ref{fig:synth-aggregation}(c). The two matrices are very similar, with only minor permutations inside the clusters.

{As in the single-layer case, through random walk dynamics we can map state nodes to points in space: indeed $p_{j\beta}(t)=\tilde{p}_l(t)$, with $l=j+\beta$ can also be seen as the component of a supra-vector in $R^{NL}$, see Appx.~\ref{appx:diffu-dist}. We then generalize Def.~\ref{def:ml-diffu-space} as follows:}

{\begin{definition}[Diffusion space - 2]
The set of state nodes $\{(i, \alpha): i = 1, \dots, N, ~\alpha = 1, \dots, L\}$ endowed with the diffusion distance at a fixed value of $t$, is a metric space. A random walk dynamics maps each state node $(i, \alpha)$ to a point in the space $\mathbb{R}^{NL}$. Furthermore, as the Euclidean norm of the probability supra-vectors, with components $p_{j\beta}(t|(i, \alpha))$, is bounded, the state nodes are embedded into a bounded subspace of $\mathbb{R}^{NL}$, which is here referred to as the \textit{diffusion space} or \textit{diffusion manifold} or \textit{diffusion embedding}.
\end{definition}}

\begin{figure*}[!htb]
\centering
\includegraphics[width=.95\textwidth]{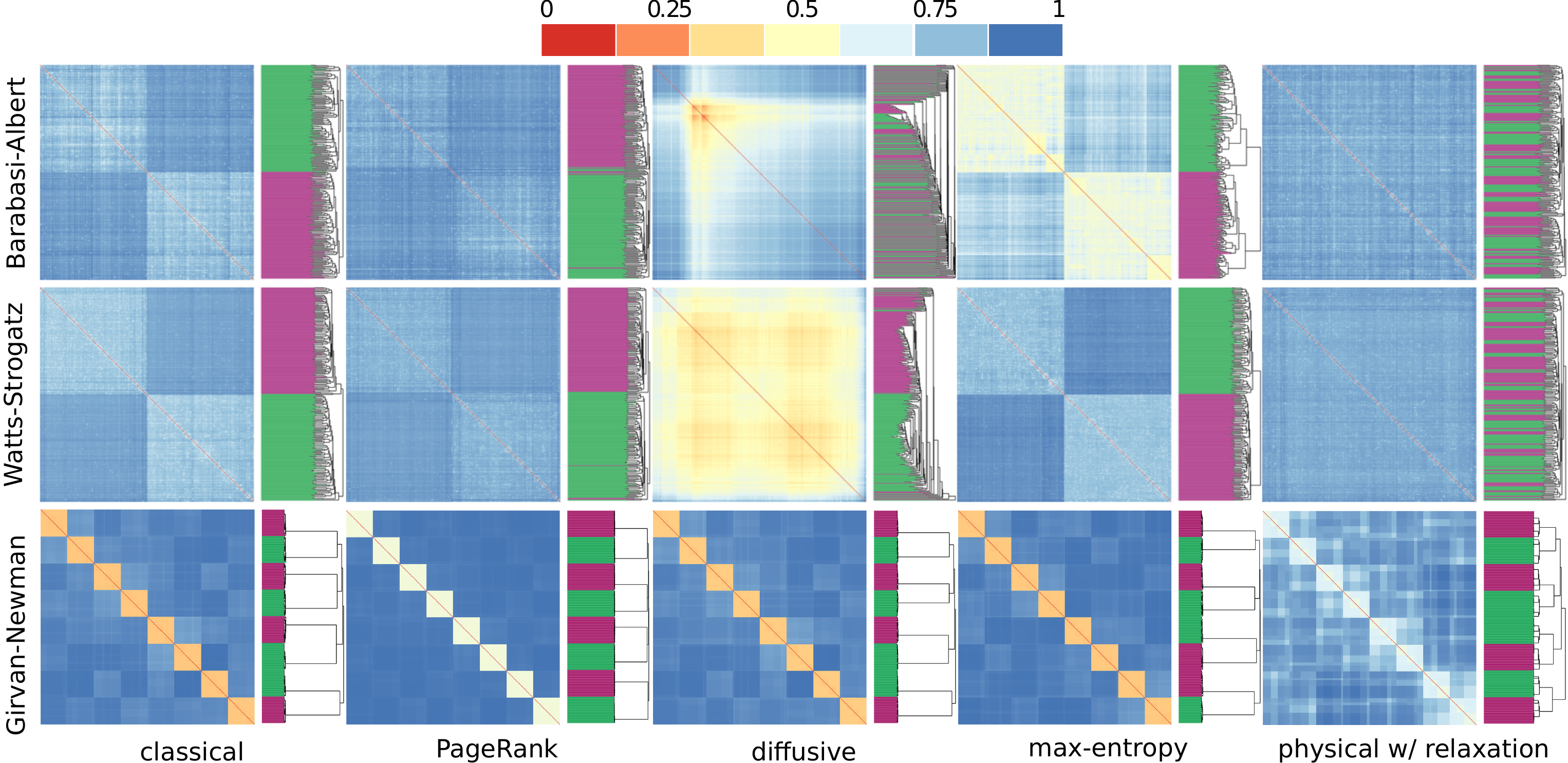}
\caption{Average diffusion distance $\bar{D}_t$ on {two-layer} multiplexes with different topologies, for fixed values of global average edge overlap (Barabasi-Albert and Watts-Strogatz) and partition overlap (Girvan-Newman) between layers. Dendrograms on the right-hand side of each distance matrix represents the corresponding hierarchical clustering to highlight the meso-scale organization of the system, with color encoding the planted node assignment in each layer. Distances have been rescaled $\frac{\bar{D}_t}{\max{\bar{D}_t}}$. The Barabasi--Albert model is characterized by the presence of hubs, which are clearly recognizable in the supra-distance matrix w.r.t. the diffusive random walk, despite the small overlap between layers. The Girvan-Newman {two-layer} multiplex has a meso-scale organized in strong communities, partly overlapping across layers. Note that at variance with edge overlapping, here partition overlapping is defined in terms of nodes belonging to the same group without requiring those nodes to be connected by an overlapping edge.}
\label{fig:overlap}
\end{figure*}

\begin{figure*}
    \centering
    \includegraphics[width=.95\textwidth]{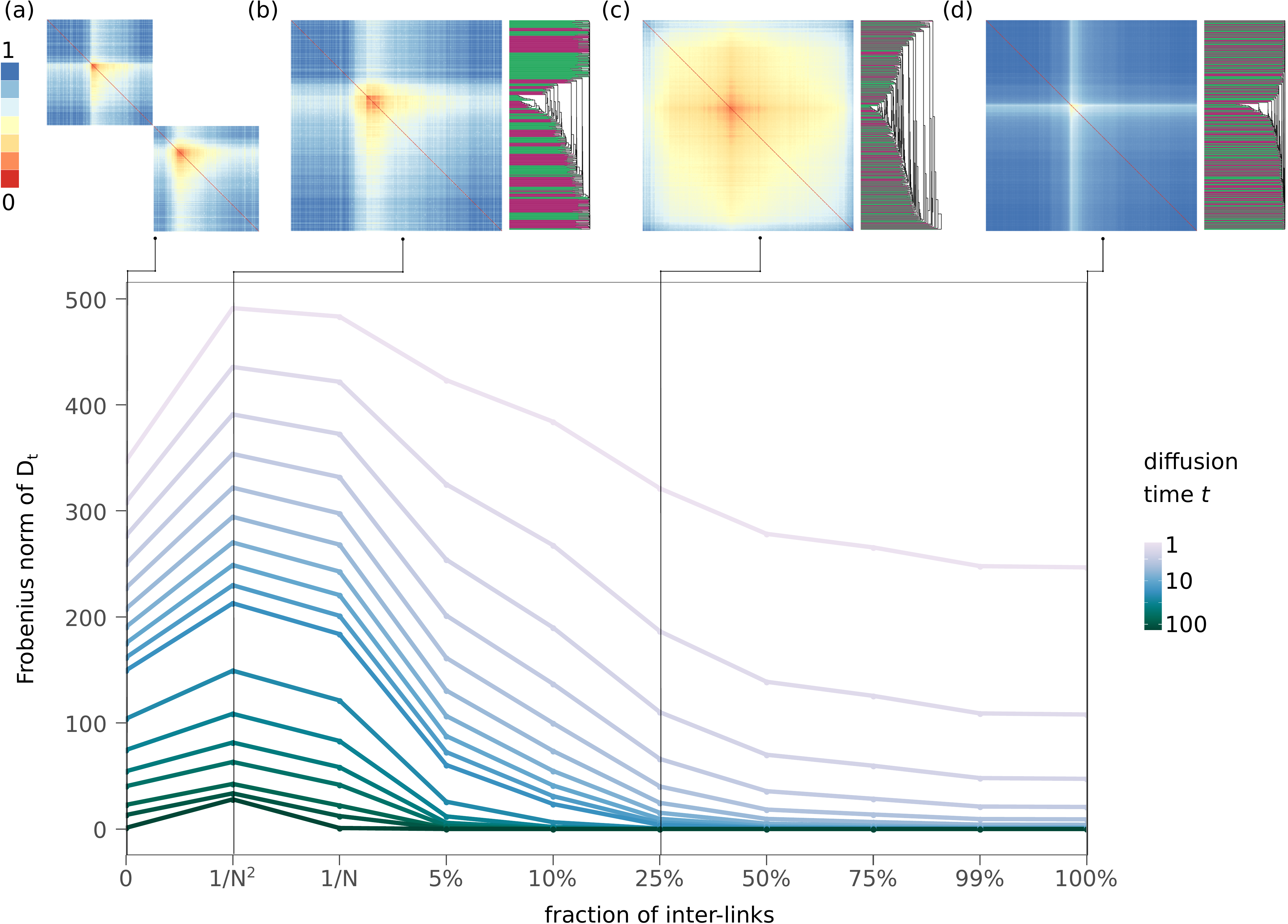}
    \caption{Frobenius norm of the supra-distance matrices of a synthetic {two-layer} network, w.r.t. a diffusive random walk. {Each layer is generated from a Barabasi-Albert model as in Fig.~\ref{fig:overlap}}. As the fraction of inter-layer links grows we move from two disconnected multiplexes to a fully inter-connected multilayer whit all $N^2$ connections across layers. The heatmaps of $D_{t=5}$ are four representatives of the different regimes: (a) uncoupled layers; (b) a single-inter-link between the replicas of a random node coupling the two layers, i.e., $\left[\frac{1}{N^2}, \frac{1}{N}\right]$ partially interconnected multiplex; [1/N] fully interconnected multiplex (all state nodes corresponding to the same physical node are interconnected); {(c)-(d)} $\left[\frac1N, 1\right]$ multilayer regime consisting of an interconnected multiplex with the addition of cross-links between state nodes of distinct physical nodes.}
    \label{fig:frobenius}
\end{figure*}

\begin{figure*}
    \centering
    \includegraphics[width=.95\textwidth]{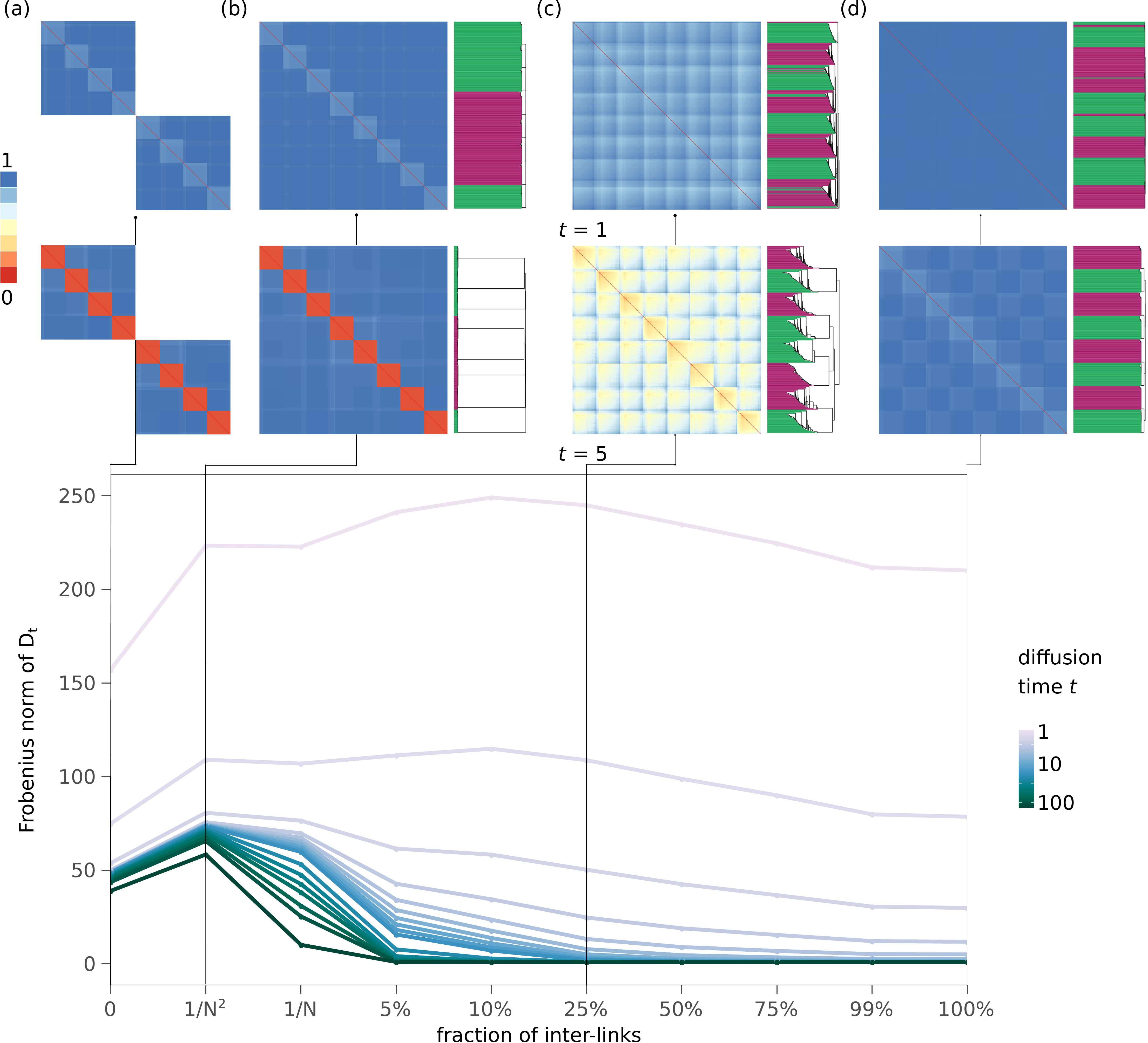}
    \caption{As in Fig.~\ref{fig:frobenius} for networks with a meso-scale organized in four strong communities in each layer and partition overlap of 1\%. The heatmaps on the top correspond to $D_{t=1}$, while the other are $D_{t=5}$.}
    \label{fig:frobenius-comm}
\end{figure*}

\subsection{Multilayer Diffusion Manifolds}

The use of different random walk dynamics to explore a system has an impact on the distances between its units and, consequently, on how the units are distributed in the induced diffusion spaces. Similarly, the diffusion time shapes the pairwise distances, highlighting local features of complex network geometry on short time scales and its more persistent structures for large diffusion times. In the multilayer setting there is an additional level of complexity given by the inter-layer connections and by the layer-layer correlations. To gain further insights, we generate 3 distinct classes of synthetic multilayer networks, with system size $N=200$, and analyze them through the lens of diffusion geometry.

\begin{figure*}[!htb]
\includegraphics[width=\textwidth]{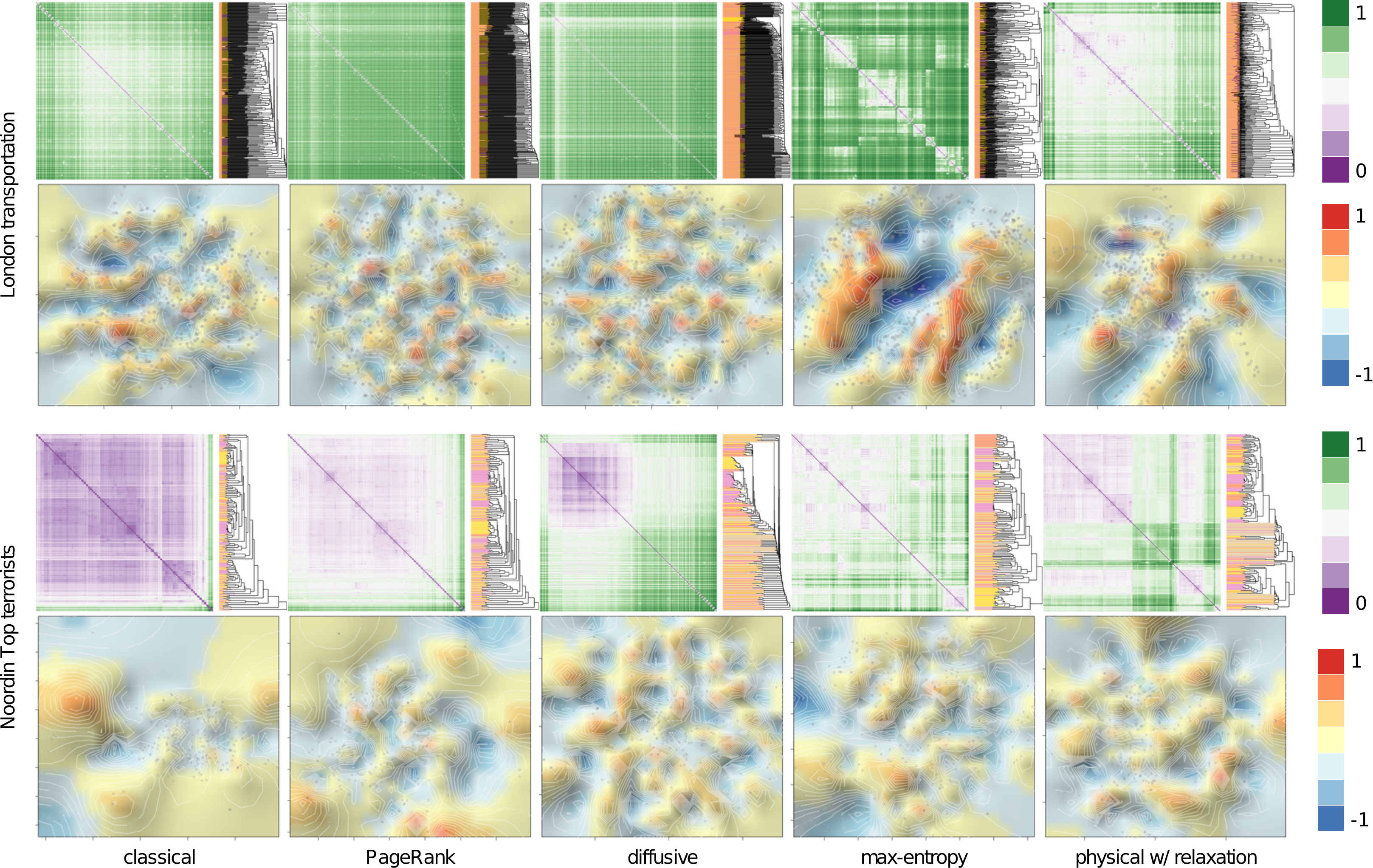}
\caption{Average diffusion distance supra-matrices $\bar{D}_t$ organized according to their hierarchical clustering as in previous analyses, and projection in $\mathbb{R}^3$ of the corresponding diffusion manifolds, in the case of the two real multiplex networks: the London public transportation network (top panels) and the social relationships of Indonesian terrorists (bottom panels). In both cases, nodes have been embedded in space, more specifically in $[-1, 1]^3$, through multidimensional scaling, a metric preserving embedding method that depends on two parameters: a distance (or dissimilarity) matrix and an embedding dimension~\cite{Cox2000} (see Appx.~\ref{appx:mds} for details). To facilitate the visualization of the three-dimensional embeddings, we draw the surface which better approximates the cloud of nodes in $\mathbb{R}^3$ and project it on the plane, encoding the third dimension with colors. Nodes are shown as dots on the top of the surfaces.}
\label{fig:manifolds}
\end{figure*}

The first class consists of Barabasi-Albert scale-free networks~\cite{barabasi1999emergence} on each layer: we consider a linear preferential attachment with 4 edges added by each new node during the growth process, while setting at 10\% the edge overlapping across layers -- defined in terms of the fraction of links which are present in both layers among the same pairs of nodes~\cite{de2015structural}.

The  second class consists of Watts-Strogatz small-world network~\cite{watts1998collective} on each layer, obtained by rewiring lattices with probability 0.2, where edge overlapping is tuned similarly to the first class.

The third class consists of layers with strong meso-scale structure organized in 4 groups, like in a Girvan-Newman model~\cite{girvan2002community} on each layer: the probability that two nodes within the same group are connected is 1, whereas cross-group connections are much sparser and present with probability 0.05.
Group overlapping~\cite{de2015structural} -- defined in terms of the fraction of nodes planted in the same group on both layers -- is fixed at 1\%.

The role of layer-layer correlations and their interplay with the distinct network topologies considered above is summarized in Fig.~\ref{fig:overlap}. As expected, there are relevant differences due to the type of random search dynamics and to the topological features of the underlying topologies.
For instance, the diffusive walk for the Barabasi-Albert system leads to a high level of mixed pathways across layers.
{In the resulting diffusion embedding the nodes are not grouped together according to their layer assignment, as can be seen from the mix of green and purple in the corresponding dendrogram of Fig.~\ref{fig:overlap}.}
For the same walk, in the case of the Watts-Strogatz system the result is the opposite: nodes aggregate into {highly homogeneous functional clusters w.r.t. the nodes layer assignment.}
As expected, a high amount of \textit{geometric mixing} {(i.e. the mixing in the embedding space of state nodes that belong to different layers or communities in the network)} is also observed when the {network functionality is studied through the} physical random walk with relaxation.
This dynamics on the Girvan-Newman model makes intra- and inter-layer distances more homogeneous than the other RWs and the cross-layer communities appear only for very large values of partition overlap, see Fig.~\ref{fig:community} in the Appx.~\ref{appx:community}.
Overall, it is not guaranteed that diffusion pathways across layers favor the geometric mixing in the diffusion manifold: the result depends on the type of dynamics and on layer-layer correlations.

{Delving deeper into this interplay between structure and dynamics}, we have considered a second battery of synthetic models, where we increasingly add inter-layer connectivity between layers.

The absence of information pathways across layers, happening for instance when two layers are not coupled together, leads naturally to disjoint diffusion manifolds, {i.e. disjoint clouds of points in the diffusion space,} each one corresponding to the distinct layers.
When the two layers are interconnected together, a trivial result is that the strength of inter-layer connectivity facilitates the flow of information across layers.
However, the above process hides an interesting phenomenon, that is unveiled in Figs.~\ref{fig:frobenius}-\ref{fig:frobenius-comm}.
To better characterize it, we calculate the Frobenius norm, which is defined as follows, for a generic matrix $\mathbf{A}$
\begin{equation*}
    \|\mathbf{A}\|_{\mathrm{F}}=\sqrt{\sum_{i=1}^{m} \sum_{j=1}^{n}\left|A_{i j}\right|^{2}}=\sqrt{\operatorname{trace}\left(\mathbf{A}^{T} \mathbf{A}\right)}
\end{equation*}
to quantify the overall intensity of an average diffusion distance matrix.
The Frobenius norm is computed when the two layers are not coupled, on the union of the two distance matrices, i.e., $\sqrt{\|\mathbf{A}\|_{\mathrm{F}}^2 + \|\mathbf{B}\|_{\mathrm{F}}^2}$, and then on the supra-distance matrices for increasing fraction of inter-layer connectivity: first state nodes corresponding to the same physical node are interconnected with each other to create an interconnected multiplex; after that this regime is reached, the cross-links between state nodes corresponding to all other physical nodes are created, until the total of $N^2$ connections is generated.
Remarkably, when one interlink is added between the layers, the Frobenius norm increases: this is due to the fact that the new link coupling the two layers creates a bottleneck for the information exchange across layers, even for large values of $t$.
Once that more inter-links are added, the Frobenius norm decreases, {tending to} a plateau when the fraction of inter-layer links approaches the 100\%.
It is worth noticing how the structures of the single layers and of the whole fully-interconnected multilayer coexist in the different diffusion spaces w.r.t. the diffusion time $t$.
In case of Barabasi-Albert networks in each monoplex, Fig.~\ref{fig:frobenius}, their typical heterogeneity {(presence of hubs)} characterizes also the diffusion manifolds. {As a matter of fact hubs have smaller diffusion distances to any other node.}
On the contrary, for small values of $t$ we can see in Fig.~\ref{fig:frobenius-comm} that the community structure is concealed by the presence of the inter-layer links.
This is totally expected, since diffusion time plays the role of a scale parameter and locally, nodes in the fully-interconnected multilayer with communities{, Fig.~\ref{fig:frobenius-comm}-(d),} are very similar.

Our results highlight that the existence of topological correlations across layers induce changes in how information is exchanged between state nodes.
Such changes alter diffusion distances and might lead to two different regimes: i) flow keeps segregated within layers and the  {diffusion manifold corresponding to the multilayer} consists of two well separated sub-manifolds representing each layer separately; ii) flow is integrated {and the diffusion pathways across layers} mix up those  sub-manifolds.

\section{Applications to empirical multilayer systems}\label{sec:applications}

We use the newly introduced family of metrics to study two real systems with multiple types of interactions: the multimodal transportation network of London~\cite{de2014navigability} and the multilayer Noordin Top terrorists network~\cite{roberts2011terrorist}.
The first system consists of three layers corresponding to the Tube, overground, and DLR, arranged in a multiplex with couplings $D(i; \alpha, \beta) = 1$ for $\alpha \neq \beta \in \{1, 2, 3\}$. Nodes represent stations ($N=369$ in total) and connections between them are weighted and undirected.
The network of interactions among 78 Indonesian terrorists ($N=79$ in the data set, but actor 58 is usually removed since it is disconnected in all layers) is a four-layers multiplex, representing their pairwise trust (T), operational (O), communication (C) ties, and business (B) relations~\cite{battiston2014}.

\begin{figure}[!ht]
    \centering
    \includegraphics[width=.45\textwidth]{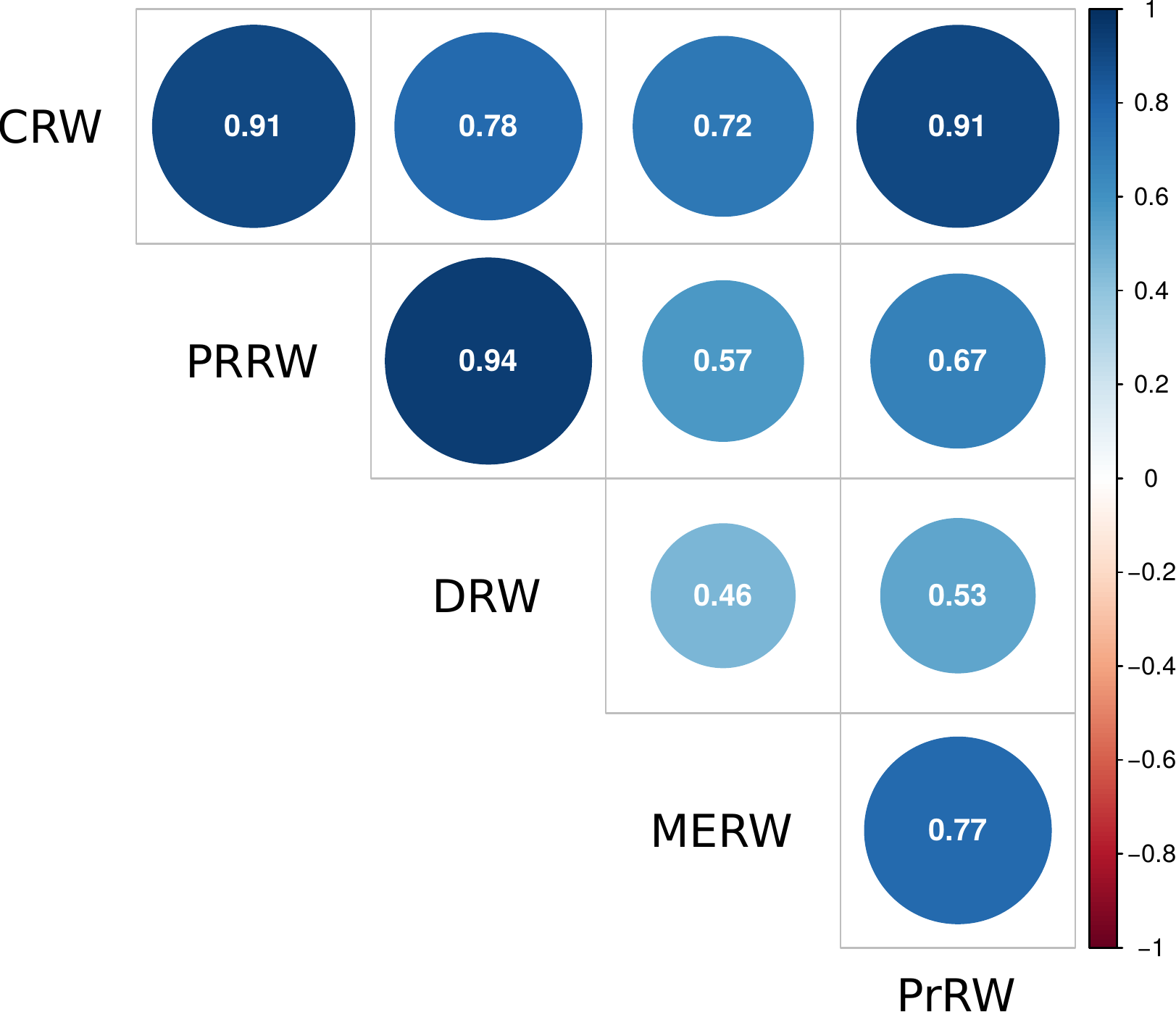}
    \caption{Estimating the similarity between diffusion manifolds corresponding to different RW dynamics, evaluated on the London multimodal transportation network. We calculate the Pearson's correlation (encoded by size and color) between the entries of pairs of supra-distance matrices (Mantel's statistic) which is then tested for significance by permutation (permutation test), with $\alpha=0.001$. The test is general, because it applies directly on distance matrices, whereas any test performed on the low-dimensional embedding of diffusion manifolds would be less precise because of the information loss during projections.}
    \label{fig:london-corr}
\end{figure}

\begin{figure}[!b]
    \centering
    \includegraphics[width=.45\textwidth]{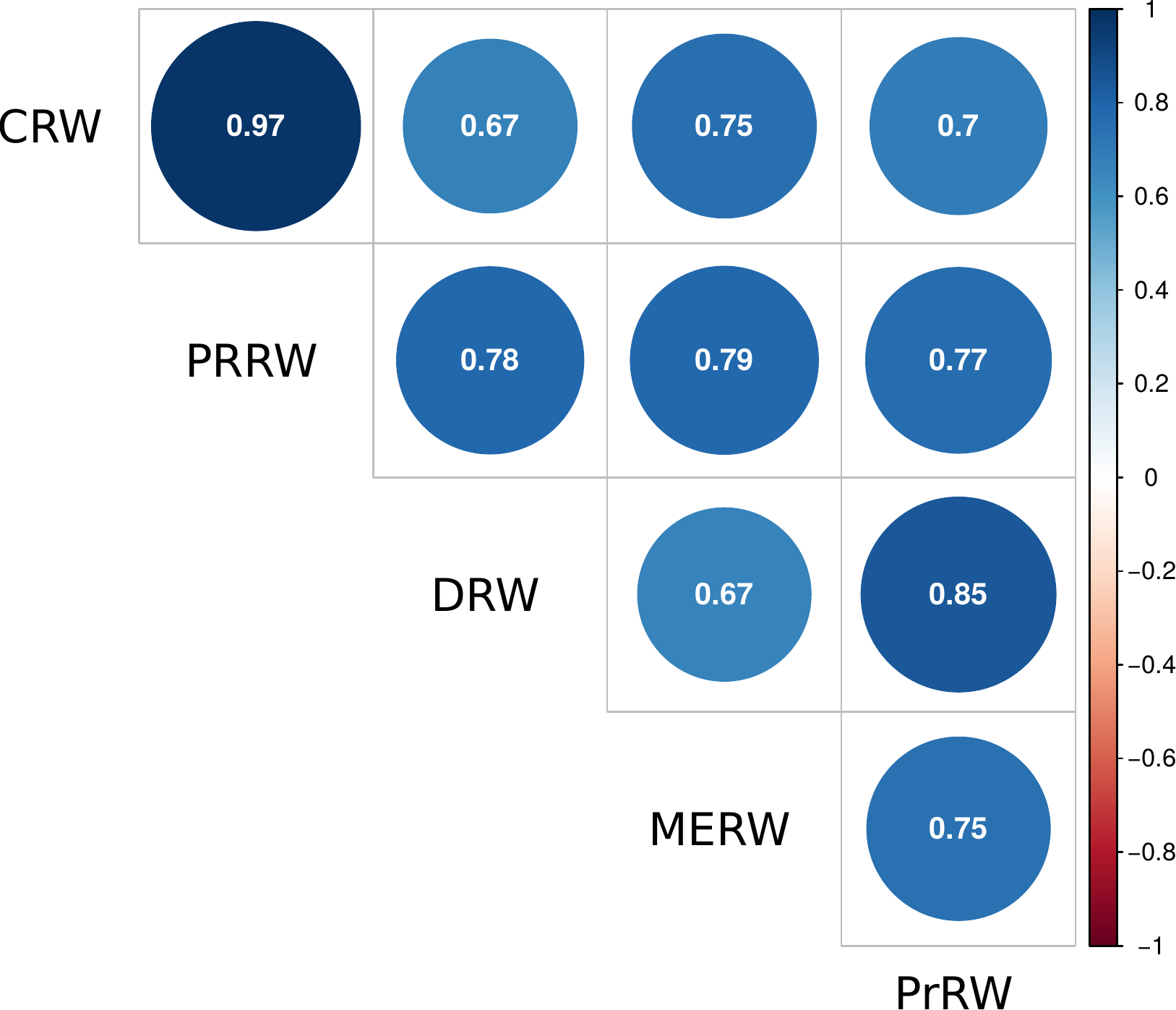}
    \caption{Same as in Fig.~\ref{fig:london-corr} for the Noordin Top terrorists network.}
    \label{fig:noordin-corr}
\end{figure}

Figure~\ref{fig:manifolds} shows the average diffusion manifolds, {i.e. the diffusion manifold corresponding to $\bar{D}_t$,} projected in $\mathbb{R}^3$ through multidimensional scaling, induced by different RW dynamics.
As observed in~\cite{de2014navigability}, the best exploration strategy, i.e. the RW to adopt to cover efficiently the network, depends on the topology of the multilayer. This is reflected in the maps of Fig.~\ref{fig:manifolds}, even though they are low-dimensional approximations of the true diffusion manifolds.
As a matter of fact, for the London transportation network, the manifolds induced by the classical, PageRank, and diffusive random walks appear qualitatively very similar with each other, and considerably different from those induced by MERW and PrRW.
Instead, the supra-distance matrices and manifolds obtained for the terrorists network appear similar in that all have a group of nodes with small pairwise distances, and another group of nodes which are distant from each other.
To quantify more adequately how diverse, or similar, the manifolds are, we compare their supra-distance matrices by means of Mantel's test~\cite{mantel1967detection,legendre2012numerical}, where the null hypothesis is that the pairwise distances in one matrix are not monotonically related to the corresponding distances in the second matrix, and show the results of our test in Figs.~\ref{fig:london-corr}-\ref{fig:noordin-corr}.

\section{Discussion and Conclusions}

We have considered different families of random walk dynamics, adequately extended to the realm of multilayer networks, to introduce the multilayer diffusion geometry.
Its classical counterpart, single-layer diffusion geometry, intimately relates metastable synchronization, consensus and random search dynamics, providing a novel framework for identifying functional clusters in complex networks.

While the framework and its validity remain the same, its natural generalization to multilayer networks -- i.e., systems consisting of multiple types of relationships among their units -- was missing.
Here, we fill this gap and provide evidence that multilayer {diffusion distance and induced manifolds} encode information due to the {both the structure} of the multilayer and the {type of dynamics} on it.
{Of course, here we focus only of averaging operators (Laplacians) and diffusion processes for studying the function of networks, while other dynamics may display different functional behaviors \cite{zhang2015explosive}. In the same spirit, different representations of complex systems accounting for other types of interactions, e.g., hypergraphs, may provide diverse insights on the functionality of complex systems \cite{de2020social}.}

From the analysis of synthetic networks with overlapping edges or groups across layers, we have found that the interplay between dynamics and topology cannot be easily decoupled: e.g., the classical random walk reveals a manifold in which state nodes of the different layers have larger distances than the intra-layer distances, but this does not remain true in the presence of strong communities, which are not necessarily overlapping.
Also the behavior of the MERW is not trivial: the top-level hierarchical structure unveiled in the Barabasi-Albert is compatible with that of the Watts-Strogatz network, despite the high heterogeneity of the first, and this could be surprising since MERW is influenced by irregularities in nodes degree.
In another scenario, where two layers are originally uncoupled and do not exchange information, we add inter-layer connectivity to better understand how the originally disjoint diffusion manifolds approach each other because of the presence of multilayer information pathways.
Our results highlight that also in the regime of partially interconnected multiplex, where not all replicas of a physical node are interacting -- which in the real world could mean a failed connection between a bus and a train station -- cross-layer pathways form, allowing the inter-layer information exchange.
Furthermore, as we move toward the fully-interconnected multilayer, distances become smaller (as shown by the Frobenius norm) and we highlighted two different behaviors of the manifolds, depending on the meso- and macro- scale of the single layers: in Barabasi-Albert model the distinctive meso-scale structure, i.e., the presence of hubs, remains clearly visible, while the community structure in the Girvan-Newman networks is washed out in the diffusion space, above all for small diffusion times.

Finally, we have applied our novel framework to two empirical multilayer systems, namely the public transportation of London and the social network of Noordin terrorists.
The diffusion geometry corresponding to different random walk dynamics are not necessarily distinct, and we have developed a quantitative method to assess the correlation between the underlying multilayer diffusion manifolds.
In the case of the transportation system, we find that the MERW, which in the synthetic networks was able to separate the layers, highlights two groups of near nodes, that are not captured by other dynamics. This may suggest that (i) the structure of this system, at different scales, has features that are characteristic of different models (e.g., heterogeneity and communities); (ii) different dynamics induce different manifolds and consequently, the analysis of networked systems embedded into space cannot exclude the analysis of the dynamics itself.
Conversely, in the case of the social system, we find that the metrics have generally higher correlations, so that their latent diffusion spaces may be likewise similar. The hierarchical structure unveiled by the supra-distance matrices seems to suggest a cross-layer core-periphery functional organization.

In this work we extended the mathematical framework for diffusion distances, introduced in~\cite{de2017diffusion}, to different random walk dynamics and different interconnected layers.
We highlighted some interesting results in the sense of functional meso-scale structures.
Nonetheless, the thorough analysis of functional multilayer communities is beyond the scope of this work and left for future work.
Our work provides a novel tool for the analysis of multilayer systems from {a} network geometry perspective~\cite{Boguna2021}.
Since the latent diffusion geometry is induced by network-driven processes, our framework provides also a complementary view to structural analysis, such as the one provided by hyperbolic network geometry~\cite{krioukov2010hyperbolic,papadopoulos2012popularity,bianconi2017emergent}, recently used to analyze multilayer networks~\cite{kleineberg2016hidden}, and higher-order analysis~\cite{battiston2020networks}.


%

\appendix

\section{Random walks and Markov chains}\label{appx:rws}

A Poisson process is a right-continuous process $(X_t)_{t \geq 0}$ with values {in} $\{0, 1, 2, \dots\}$ and holding times $S_1, S_2, \dots$ ($S_i = J_i - J_{i-1}$ is the time occurring between the random jump times $J_{i-1}$ and $J_i$) that are independent exponential random variables of rate $0 < \lambda < \infty.$
In a generalized Poisson process (or birth process) the parameter $\lambda$ is allowed to depend on the current state of the process.
Given its \textit{birth rates} $0 \leq q_i < \infty$ for $i = 0, 1, 2, \dots$, $(S_i)_{i\in \mathbf{N}^+}$ are independent exponential random variables with rates $q_i$.
Finally, a continuous-time Markov chain (MC) $(X_t)_{t\geq 0}$ on a finite set $I$ with generator $Q$ and initial distribution $\mathbf{p}_0$ can be described in terms of a Poisson process.
Each state $i \in I$ of the process is a chamber and doors close the passage to the other states.
From time to time a single door opens (events cannot be simultaneous) allowing the process to change state and
the doors open at the jump times of a Poisson process of rate $q_{ij}$~\cite{norris1998markov}.
The \textit{generator} of the MC is indicated by $Q$, because it is a particular matrix, called $Q-$matrix in~\cite{norris1998markov}, satisfying three conditions
\begin{itemize}
  \item[(i)] $0 \leq -q_{ii} < \infty$
  \item[(ii)] $q_{ij} \geq 0 \quad \forall i \neq j$
  \item[(iii)] $\sum\limits_{j} q_{ij} = 0 \quad \forall i$.
\end{itemize}
To recap, a continuous-time MC can be (equivalently) defined in terms of its jump chain and holding times, or of its transition probabilities given by the solution of the forward equation~\cite[Thm.2.8.2]{norris1998markov}.

\section{On the diffusion distance(s) of single-layer networks}\label{appx:diffu-dist}

{In this appendix we assume that $G=(V, E)$ is an undirected, weighted, and connected network and we focus on the (single-layer) diffusion distance based on the classical random walk.
In this case, we have that the random walk normalized Laplacian $\mathbf{\tilde{L}}=\mathbf{I} - \mathbf{D}^{-1}\mathbf{W}=\mathbf{D}^{-\frac12}\mathcal{L}\mathbf{D}^{\frac12}$, where $\mathcal{L}^{\text{sym}} = \mathbf{D}^{-\frac12}(\mathbf{D} - \mathbf{W})\mathbf{D}^{-\frac12}$ is the the symmetric normalized Laplacian matrix \cite{chung1997spectral}.
The latter has a spectrum of real, non-negative eigenvalues and its eigenvectors can be chosen to form an orthogonal matrix $\mathbf{Q}$, so that $\mathcal{L}^{\text{sym}} = \mathbf{Q} \boldsymbol{\Lambda} \mathbf{Q}^T$.
It follows that $\tilde{\mathbf{L}} = \mathbf{D}^{-\frac12} \mathbf{Q} \Lambda \mathbf{Q}^T \mathbf{D}^{\frac12}$ and consequently $e^{-t \tilde{\mathbf{L}}} = \mathbf{D}^{-\frac12} \mathbf{Q} e^{-t\boldsymbol{\Lambda}} \mathbf{Q}^T \mathbf{D}^{\frac12}$.
Let us indicate by $\phi_{\ell}(i)=Q^i_j$ the $i-$th component of the $\ell-$th eigenvector of $\mathcal{L}^{\text{sym}}$, then
\begin{align*}
    \left( e^{-t \tilde{\mathbf{L}}} \right)^i_k & =  \sum_{\ell=1}^N \frac{\phi^{\ell}(i)}{\sqrt{s_i}} \phi_{\ell}(k)\sqrt{s_k}e^{-t\lambda_{\ell}} \\
    & = \sum_{\ell=1}^N \psi^{\ell}(i) \varphi_{\ell}(k) e^{-t\lambda_{\ell}}
\end{align*}
with $\psi^{\ell}(i) = \frac{\phi^{\ell}(i)}{\sqrt{s_i}}$ and $\varphi_{\ell}(k)=\phi_{\ell}(k)\sqrt{s_k}$ being respectively the $i-$th and $k-$th components of the $\ell-$th right and left eigenvectors of $\tilde{\mathbf{L}}$.
We can write the square diffusion distance at a fixed time $t>0$ between $i, j \in V$ (defined in Eq.~\eqref{eq:diffu-dist-monoplex}) as
\begin{align}
    D^2_t(i, j) = \sum_{k=1}^N & \left[ \left( e^{-t \tilde{\mathbf{L}}} \right)^i_k - \left( e^{-t \tilde{\mathbf{L}}} \right)^j_k \right]^2 \nonumber\\
    = \sum_{k=1}^N & \Bigg[ \sum_{\ell=1}^N \frac{\phi^{\ell}(i)}{\sqrt{s_i}} \phi_{\ell}(k)\sqrt{s_k}e^{-t\lambda_{\ell}} \nonumber \\
    & ~ - \sum_{\ell=1}^N \frac{\phi^{\ell}(j)}{\sqrt{s_j}} \phi_{\ell}(k)\sqrt{s_k}e^{-t\lambda_{\ell}} \Bigg]^2 \nonumber \\
    = \sum_{k=1}^N & s_k \sum_{\ell=1}^N e^{-2t\lambda_{\ell}} \left( \frac{\phi^{\ell}(i)}{\sqrt{s_i}} - \frac{\phi^{\ell}(j)}{\sqrt{s_j}} \right)^2. \label{eq:spectral-diffu-dist-mono}
\end{align}
Now, exploiting the spectral representation of the exponential Laplacian matrix, we show that for a fixed value of $t$, $D_t$ is a distance function.
Here, we assume that $p^i_t(t) = \left( e^{-t \tilde{\mathbf{L}}} \right)^i_j \neq \pi_j$, that is if and invariant distribution $\boldsymbol{\pi}$ for the Markov chain exists, $t$ is small enough not to be in equilibrium.
Firstly, for any pair of nodes $i, j$, $D_t(i, j)=D_t(j, i)$ and $D_t(i, j)\geq 0$.
$D_t(i, j) = 0$ if $i=j$.
The case $\left( e^{-t \tilde{\mathbf{L}}} \right)^i_k = \left( e^{-t \tilde{\mathbf{L}}} \right)^j_k$ for all $k$ is impossible, since it would imply two rows of $e^{-t \tilde{\mathbf{L}}}$ be identical.
This, in turn, would imply $\det{\left( e^{-t \tilde{\mathbf{L}}} \right)}=0$, but $\det{\left( e^{-t \tilde{\mathbf{L}}} \right)}=\det{\left( e^{-t \boldsymbol{\Lambda}} \right)}=\prod_{\ell} e^{-t \lambda_{\ell}} > 0$.
Hence, $D_t(i, j) = 0$ if and only if $i=j$.
The triangular inequality follows from the triangular inequality of the norm norm.
Finally, $D_t$ is an Euclidean distance~\cite{liberti2014} and the diffusion distance matrix is an Euclidean distance matrix.
As a matter of facts, $D_t$ is the Euclidean distance in $\mathbb{R}^N$ between the row vectors forming $e^{-t \tilde{\mathbf{L}}}$.}

{The physical interpretation of the diffusion distance can be found looking at Eq.~\eqref{eq:diffu-dist-monoplex} written as
\begin{align*}
    D_t^2(i, j)&= \norm{\mathbf{p}(t | i)}_2^2 +
                  \norm{\mathbf{p}(t | j)}_2^2 -
                  2 \langle \mathbf{p}(t | i), \mathbf{p}(t | j) \rangle\\
                &=\sum_k \left(p_k(t | i)^2 + p_k(t | j)^2 - 2 p_k(t | i)p_k(t | j)\right)
\end{align*}
The square terms correspond to the joint probability of two random walkers starting both in the same node to be found at node $k$ at time $t$ (i.e., two independent copies of the same random process), while $p_k(t | i)p_k(t | j)$ corresponds to the probability that they meet in $k$, starting one in $i$ and the other in $j$.
Since $D_t^2(i, j)\geq 0$ we have that $\norm{\mathbf{p}(t | i)}_2^2 + \norm{\mathbf{p}(t | j)}_2^2 \geq 2 \langle \mathbf{p}(t | i), \mathbf{p}(t | j) \rangle$, i.e., given a pair of random walkers, the probability that they meet at the same node at time $t$ if they start both at $i$ or both at $j$ is larger than if they start one at $i$ and one at $j$.
If there are many walks connecting $i$ and $j$ then $\sum_k p_k(t | i)p_k(t | j)$ is large and, consequently, $D_t(i, j)$ becomes small.
we can see that this distance is small if $\langle \mathbf{p}(t | i), \mathbf{p}(t | j) \rangle = \sum_k p_k(t | i)p_k(t | j)$ is large.}

{Coifman and Lafon, in their paper~\cite{coifman2006diffusion}, suggested another interpretation of $D_t$ (based on discrete-time Markov chains) as a distance between bump functions.
The authors observed that for a fixed $x$ in a data set $X$, the transition probability out of $x$ at time $t$, $p_t(x, \cdot)$, is a bump function centered at $x$ and of width increasing with $t$.
In network terms, $p^i_{\cdot}(t)$ is a function defined on $V$ and centered at $i$; as time increases the random walker explores a larger part of the network and there is a larger probability to find it far away from the starting node $i$.}

\begin{figure*}[!htb]
\centering
\includegraphics[width=.75\textwidth]{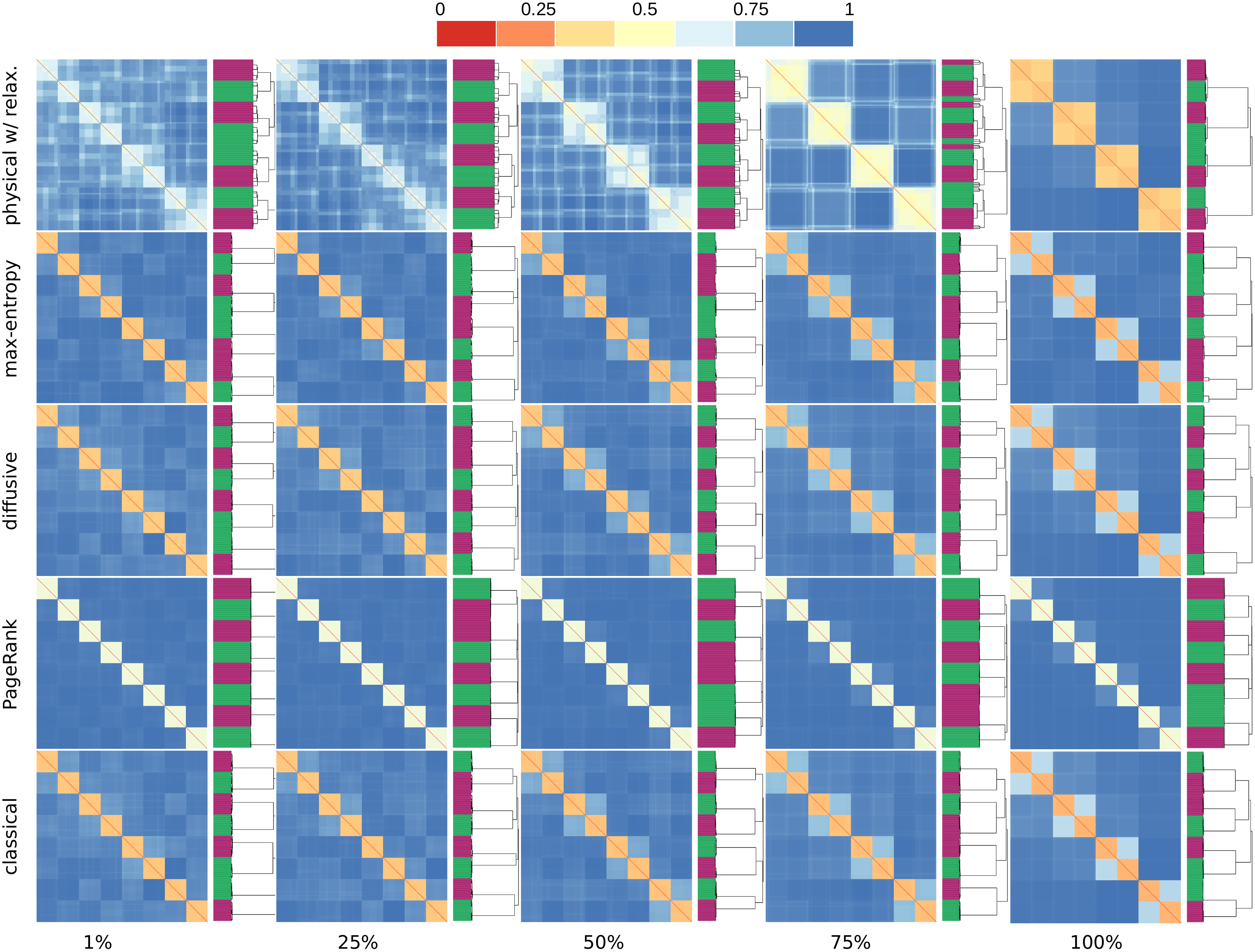}
\caption{Average diffusion distance $\bar{D}_t$ for a {two-layer} multiplex with meso-scale organization in strong communities, partly overlapping across layers.
Note that at variance with edge overlapping, here partition overlapping is defined in terms of nodes belonging to the same group without requiring those nodes to be connected by an overlapping edge.}
\label{fig:community}
\end{figure*}

\section{Multidimensional scaling}\label{appx:mds}

{Multidimensional scaling (MDS) is a collection of methods aiming to find structure in high-dimensional data, given a $n \times n$ dissimilarity matrix $\Delta$ between the observations and the dimension $p$ of the Euclidean space in which data are embedded~\cite{Cox2000}.
The final goal is to find a space configuration of the data points in $\mathbb{R}^p \ni \mathbf{x}_1, ..., \mathbf{x}_n$, such that $\norm{\mathbf{x}_i - \mathbf{x}_j} \approx \delta_{ij}$, where $\delta_{ij}$ are the original distances and $\norm{\cdot}$ is the Euclidean distance in $\mathbb{R}^p$.}

{The MDS problem can be solved analytically or using iterative procedures, like the majorisation algorithm.
The first case, involves a double-centering step of the dissimilarity matrix and then its eigendecomposition.
The second, involves the minimization of a loss function, in particular, Kruskal's \textit{stress-1}.
Given a configuration $X$ of $N$ points in $\mathbb{R}^p$ with $1 \leq p \leq N-1$
\begin{equation}\label{eq:stress-1}
  \sigma(X) = \sum_i \sum_j \left( \delta_{ij} - d_{ij}(X) \right)^2
\end{equation}
If $\Delta$ is not a proper distance matrix, $\delta_{ij}$ are called dissimilarities and they may need to be transformed into $\hat{d}_{ij}=f(\delta_{ij})$ before the scaling, for instance through a strictly increasing monotonic function $f$ such that order is preserved $\delta_{ij} < \delta_{kl} \Rightarrow f(\delta_{ij}) < f(\delta_{kl})$.}

{In this work we use the second approach (MDS solved through majorization) applied to distance or supra-distance matrices and with embedding dimension $p=3$.
The R package and function used is \textbf{smacof::mds}~\cite{deLeeuw2009}.}

\section{Diffusion distance and community overlap}\label{appx:community}

Figure~\ref{fig:community} shows how the diffusion geometry of networks with strong meso-scale organization is influenced by layer-layer correlations, in particular community overlap.
As the overlap between the communities in the two layers grows, macro-communities across layers emerge.

\end{document}